\documentclass{emulateapj}

\usepackage{color}

\slugcomment{}

\shorttitle{The core--cusp problem in cold dark matter halos}
\shortauthors{Ogiya \& Mori}

\begin{document}


\title{The core--cusp problem in cold dark matter halos and supernova feedback: effects of oscillation}


\author{Go Ogiya and Masao Mori}
\affil{Center for Computational Sciences, University of Tsukuba, 1-1-1 Tennodai, Tsukuba 305-8577, Japan}
\email{ogiya@ccs.tsukuba.ac.jp}

%
%
%
%

\begin{abstract}
This study investigates the dynamical response of dark matter (DM) halos to recurrent starbursts in forming 
less-massive galaxies to solve the core-cusp problem. The gas, which is heated by supernova feedback after a starburst,  
expands and the star formation then terminates. This expanding gas loses energy by radiative cooling and 
then falls back toward the galactic center. Subsequently, the starburst is enhanced again. This cycle of 
expansion and contraction of the interstellar gas leads to a repetitive change in the gravitational potential 
of the gas. The resonance between DM particles and the density wave excited by the oscillating potential plays 
a key role in understanding the physical mechanism of the cusp-core transition of DM halos. DM halos effectively 
gain kinetic energy from the baryon potential through the energy transfer driven by the resonance between 
the particles and density waves. We determine that the critical condition for the cusp-core transition is such 
that the oscillation period of the gas potential is approximately the same as the local dynamical time of DM halos. 
We present the resultant core radius of a DM halo after the cusp-core transition induced by the resonance  
by using the conventional mass density profile predicted by the cold dark matter models. 
Moreover, we verify the analytical model by using $N$-body simulations, and the results validate the resonance model.
\end{abstract}


%
%
%
%

\section{INTRODUCTION}
The cold dark matter (CDM) cosmology, which is the standard paradigm of structure formation in the universe, contains 
several important unsolved problems. Recent observation of nearby less-massive galaxies and low-surface-brightness 
galaxies has revealed that the density profile of a dark matter (DM) halo is constant at the center of these galaxies 
(e.g., Moore 1994; Burkert 1995; de Blok et al. 2001; Swaters et al. 2003; Weldrake et al. 2003; Gentile et al. 2004; 
Spekkens et al. 2005; Kuzio de Naray et al. 2006, 2008; Oh et al. 2008, 2011). However, in cosmological $N$-body 
simulations based on collisionless CDM, DM halos always exhibit steep power-law mass density distribution at their
centers (e.g., Navarro et al. 1997; Fukushige \& Makino 1997; Moore et al. 1998, 1999; Jing \& Suto 2000; 
Klypin et al. 2001; Fukushige et al. 2004; Navarro et al. 2004; Diemand et al. 2004; Reed et al. 2005; 
Stadel et al. 2009; Navarro et al. 2010; Ishiyama et al. 2013).
This discrepancy is a well-known unresolved issue in the CDM scenario, and is known as the ``core--cusp problem''.

To solve this problem, our study focuses on the manner in which baryon components gravitationally affect DM halos, 
because less-massive galaxies, such as dwarf galaxies, are more sensitive to stellar  activity such as supernova 
explosions compared to giant galaxies, owing to their shallow gravitational potential. The hydrodynamic 
response of dwarf galaxies to stellar activity depends on various factors such as their mass, morphology, and star 
formation rate (e.g., Dekel \& Silk 1986; Yoshii \& Arimoto 1987; Mori et al. 1997, 1999, 2002; MacLow \& Ferrara 1999; 
Silich \& Tenorio-Tagle 2001; Bland-Hawthorn et al. 2011).

In less-massive galaxies with smooth gas distribution, supernova feedback blows the interstellar gas out from the galaxy 
centers. This mass loss reduces the depth of the gravitational potential around the centers of DM halos and may flatten 
their central cusps. The dynamical effects of this process on DM halos have been studied using collisionless $N$-body 
simulations (Navarro et al. 1996a; Gnedin \& Zhao 2002; Read \& Gilmore 2005; Ogiya \& Mori 2011; Ragone-Figueroa et al. 2012). 
Ogiya \& Mori (2011) discussed the relationship between the dynamical responses of DM halos and the timescale of mass 
loss (i.e., star formation rate). They found that the density profiles of DM halos correlate with the mass-loss timescale, 
and are flatter when the mass loss occurs over a short period of time than when it occurs over a long period of time. However, even if the mass loss 
occurs instantaneously, which results in a maximal effect, the central cusp remains; thus, Ogiya \& Mori (2011) concluded 
that mass-loss is not a prime mechanism in the flattening of the central cusp. Moreover, they found that the density profile 
of the DM halo recovers its initial cusp profile when the mass loss timescale is significantly longer than the dynamical 
time of the system.

Massive galaxies with moderate stellar activity are more robust against supernova feedback; therefore, supernovae can 
expel only a small fraction of the gas from such galaxies. The gas, which is heated by supernova feedback shortly after a star 
formation burst, expands temporarily and the star formation is then terminated. The expanding gas loses energy by radiative 
cooling and then falls back toward the galactic center. Then, the star formation is re-enhanced and subsequently ignites 
a starburst (Stinson et al. 2007). This repetitive gas motion leads to a repeated change in the gravitational potential. 
Previous studies have used $N$-body simulations to study the manner in which such recurring changes in the gravitational 
potential affect the density structures of DM halos (e.g., Mashchenko et al. 2006). By using hydrodynamic simulations, 
combined with star formation and supernova feedback, the dynamical effects of such recurring changes have been studied 
(Mashchenko et al. 2008; Governato et al. 2010; Pontzen \& Governato 2012; Macci{\`o} et al. 2012; Teyssier et al. 2013), 
and it has been demonstrated that oscillations in the gravitational potential may modify the central cusp into a flat core. 
Read \& Gilmore (2005) reported the importance of recurring changes in the gravitational potential. The oscillation 
timescale is also an important factor in determining the dynamical effects on DM halos, and dwarf galaxies in the Local 
Group each have distinctive star formation histories (SFHs; e.g., Mateo 1998; Tolstoy et al. 2009; McQuinn et al. 2010a, 
2010b; Weisz et al. 2011). Furthermore, recurring star formation activities duplicate the properties of dwarf spheroidal galaxies, 
which include metal distribution and the velocity-dispersion profile (Ikuta \& Arimoto 2002; Marcolini et al. 2006, 2008). 
However, several additional mechanisms have been proposed to solve the core-cusp problem. Representative models involve 
the dynamical friction induced by the motion of gas or stellar clumps (e.g., Inoue \& Saitoh 2011) and bar instability 
(Weinberg \& Katz 2002). Del Popolo (2009) discussed the composite effects of these mechanisms on the density profiles 
of DM halos.

In the present study, we investigate the physical mechanism of the cusp--core transition under recurring changes in the 
gravitational potential of interstellar gas. As previously described, hydrodynamic simulations have demonstrated the 
repetition of large-scale outflows and inflows. However, the model of Mashchenko et al. (2006) does not reflect such 
phenomena; they used three particles with motion similar to that of harmonic oscillators to express the repetitive 
potential change in their standard runs. 
Pontzen \& Governato (2012) constructed an analytical model to understand the mechanism through which the recurring 
potential flattens the cusps of DM halos. Their model assumes that a harmonic oscillator has the potential to govern the motion 
of the test particle and that the spring constant changes instantaneously and adiabatically. They determined that most 
particles in the system gain energy when the potential change occurs instantaneously, and thus the systems expand. However, 
the nature of the harmonic oscillator potential differs significantly from that of the gravitational potential, such that 
the acceleration of the harmonic oscillator increases with $r$, but gravity decays. 

To determine the fundamental physical 
mechanism of the cusp--core transition of DM halos, we adopt a more reasonable formula for the periodic potential by using 
the Fourier expansion, and analyze the dynamical response of DM halos to recurring potential change with arbitrary periods. 
Our model can precisely predict the resultant core scale. In addition, we evaluate the energy transfer rate from the oscillating 
potential to the DM halo, and estimate the number of oscillation cycles required to flatten the cusp and core scale. 
Then, we clarify the cusp-core transition of DM halos by using collisionless $N$-body simulations. The rest 
of this paper is structured as follows. Our analytic resonance model is described in Section 2. In Section 3, we present our 
numerical simulation results and interpret them using the resonance model. 
In Section 4, we discuss the energy transfer rate and some implications for the resonance mechanism. 
Finally, we summarize the results in Section 5.

%
%
%
%

\section{ANALYTIC MODEL}
We construct a simple analytic model to examine the energy transfer between DM particles and the density waves induced 
by the recurring expansions and contractions of interstellar gas. In this case, the baryon acts as a forcing potential, 
and the resonance theory is applicable to the analysis of the energy transfer from the baryon to the DM halo. 
Although the prescription in this section may appear to be somewhat ad hoc and may raise some quantitative uncertainties, it is designed to provide an intuitive understanding of the physical mechanism responsible for the cusp--core transition and gives a useful formula for predicting the resultant core scale created by the oscillatory external force. 
Section 2.1 describes the resonance model under an ideal situation, and Section 2.2 shows the resonance condition. The simple 
relationship between the core radius of a DM halo and the oscillation period is derived in Section 2.3.


\subsection{Resonance Model}
In this section, the dynamical evolution of a DM halo is assumed to be described by the linearized fluid equations of 
the perturbed equilibrium state. 
Hereafter, we label physical quantities in the initial equilibrium state as ``0'', 
the external forces as ``ex'', and the induced quantities as ``ind''. We assume an infinite constant density field as 
the equilibrium state ($\rho_{\rm 0} = const.$) and select a group of particles with constant velocity  ($v_{\rm 0} = const.$).
In addition, the absolute values of the induced quantities are assumed to be significantly smaller than those of 
the equilibrium state and the external force. With these assumptions, the linearized equation of motion is given by
\begin{equation}
\frac{\partial v_{\rm ind}(t,r)}{\partial t } + v_{\rm 0} \frac{\partial v_{\rm ind}(t,r)}{\partial r} = - \frac{\partial \Phi_{\rm ex}(t,r)}{\partial r}, \label{Euler}
\end{equation}
where the induced potential, $\Phi_{\rm ind}$, is neglected.

By using the Fourier expansion,  the periodic external force is expressed as 
\begin{equation}
-\frac{\partial \Phi_{\rm ex}(t,r)}{\partial r} = \sum_{n=1}^{\infty} A_n \cos{(kr - n \Omega t)},
\end{equation}
where $A, k$, and $\Omega$ represent the oscillation strength, the wavenumber,
and the angular frequency of the external force, respectively. The index, $n$, corresponds to the $n$th overtone mode. 
The solution of Equation (\ref{Euler}) is given by 
\begin{eqnarray}
v_{\rm ind}(t,r) &=& \sum_{n=1}^{\infty} v_{{\rm ind,}n}(t,r) \nonumber \\
                &=& -\sum_{n=1}^{\infty} \frac{A_n}{n \Omega - k v_{\rm 0}} \nonumber \\ 
                &\times& \{ \sin{(kr - n \Omega t)} - \sin{(kr - kv_{\rm 0}t)} \}. \label{v_ind}
\end{eqnarray}

The linearized equation of continuity is similarly obtained:
\begin{equation}
\frac{\partial \rho_{\rm ind}(t,r)}{\partial t} + v_{\rm 0} \frac{\partial \rho_{\rm ind}(t,r)}{\partial r} 
= - \rho_{\rm 0} \frac{\partial v_{\rm ind}(t,r)}{\partial r}, \label{continuity}
\end{equation}
and the solution is given by 
\begin{eqnarray}
\rho_{\rm ind}(t,r) &=&  \sum_{n=1}^{\infty} \rho_{{\rm ind,}n}(t,r) \nonumber \\
                   &=& -\sum_{n=1}^{\infty} \frac{A_n \rho_{\rm 0} k}{(n \Omega - k v_{\rm 0})^2} \{ \sin{(kr - n \Omega t)} - \sin{(kr - kv_{\rm 0}t)} \nonumber \\
                   &+& (n \Omega - k v_{\rm 0})t \cos{(kr - kv_{\rm 0}t)} \}. \label{rho_ind}
\end{eqnarray}

These solutions satisfy the reasonable initial conditions $v_{\rm ind}(0,r)=0$ and $\rho_{\rm ind}(0,r)=0$.
Although the coefficients of $v_{{\rm ind,}n}$ and $\rho_{{\rm ind,}n}$  appear to diverge when $n \Omega$  approaches $k v_{\rm 0}$, 
we can derive finite values for this limit by using ${\rm l'H\hat{o}pital's}$ rule:
\begin{eqnarray}
\lim_{n \Omega \rightarrow k v_{\rm 0}} v_{{\rm ind,}n}(t,r) = A_n t \cos{(kr - kv_0 t)}, \label{res_v} 
\end{eqnarray}
and
\begin{eqnarray}
\lim_{n \Omega \rightarrow k v_{\rm 0}} \rho_{{\rm ind,}n}(t,r) = \frac{A_n \rho_0 k}{2} t^2 \sin{(kr - kv_0 t)}. \label{res_rho}
\end{eqnarray}
Equations (\ref{res_v}) and (\ref{res_rho}) clearly show that the effects of resonances increase with time. 

Next, we examine the energy interchange between the DM collisionless system and the oscillatory external force on the basis 
of resonance. By averaging the external force per volume $F = -(\rho_{\rm 0} + \rho_{\rm ind}) {\bf \nabla} \Phi_{\rm ex}$ over 
a wavelength, the term of $\rho_{\rm 0}$ vanishes because of the nature of the trigonometric function:
\begin{eqnarray}
<F> &=& < \sum_{m=1}^{\infty} A_m \cos{(kr - m \Omega t)} \nonumber \\
    &\times& \sum_{n=1}^{\infty} \frac{- A_n \rho_{\rm 0} k}{(n \Omega - k v_{\rm 0})^2} [\sin{(kr -n  \Omega t)} - \sin{(kr - kv_{\rm 0}t)} \nonumber \\
    &+& (n \Omega - k v_{\rm 0})t \cos{(kr - kv_{\rm 0}t)}] >.
\end{eqnarray}
Here, the interactions between different Fourier modes (i.e., $m \neq n$) are neglected, and only the self-interaction of 
the modes (i.e., $m = n$) is considered. Spatially averaged force is determined through arithmetic calculations:
\begin{eqnarray}
<F> &=& \sum_{n=1}^{\infty} \frac{A_n^2 \rho_{\rm 0} k}{2 (n \Omega - k v_{\rm 0})^2} \nonumber \\
    &\times& \{ \sin{[(n \Omega - k v_{\rm 0})t]} - (n \Omega - k v_{\rm 0})t \cos{[(n \Omega - k v_{\rm 0})t]} \} \nonumber \\
                    &=& \sum_{n=1}^{\infty} \frac{A_n^2 \rho_{\rm 0}}{2} \frac{d}{d v_{\rm 0}} \biggl \{ \frac{\sin{[(n \Omega - k v_{\rm 0})t]}}{n \Omega - k v_{\rm 0}} \biggr \} . \label{ave_f}
\end{eqnarray}

Note that we have thus far confined the argument to a particular group of particles. In actuality, the system consists of 
uncounted groups of particles and each group has a respective velocity in the equilibrium state. The energy transfer rate 
for the entire system, $\frac{d K}{dt}$, is estimated as
\begin{equation}
\frac{d K}{dt} = \int_{-\infty}^{\infty} d^3 v_{\rm 0} <F> v_{\rm 0} f(v_{\rm 0}), \label{dkdt}
\end{equation}
where $f(v_{\rm 0})$ is the distribution function (DF) of velocity. Assuming isotropic velocity field and applying Equation (\ref{ave_f}), 
Equation (\ref{dkdt}) becomes
\begin{eqnarray}
  \frac{d K}{dt} &=& 4 \pi \sum_{n=1}^{\infty} \frac{A_n^2 \rho_{\rm 0}}{2} \int_{0}^{\infty} d v_{\rm 0} f(v_{\rm 0}) v_{\rm 0}^3 \frac{d}{d v_{\rm 0}} \biggl \{ \frac{\sin{[(n \Omega - k v_{\rm 0})t]}}{n \Omega - k v_{\rm 0}} \biggr \} \nonumber \\
                 &=& 2 \pi \sum_{n=1}^{\infty} A_n^2 \rho_{\rm 0} \biggl \{ f(v_{\rm 0}) v_{\rm 0}^3 \frac{\sin{[(n \Omega - k v_{\rm 0})t]}}{n \Omega - k v_{\rm 0}} \biggr |_{0}^{\infty} \nonumber \\
                 &-& \int_{0}^{\infty} d v_{\rm 0} \frac{d}{d v_{\rm 0}} [f(v_{\rm 0}) v_{\rm 0}^3] \frac{\sin{[(n \Omega - k v_{\rm 0})t]}}{n \Omega - k v_{\rm 0}} \biggr \}. \label{partial_integration}
\end{eqnarray}
Here, we have used partial integration, and the first term of the second row is 0 because $f(v_{\rm 0})$ dumps faster than 
$v_{\rm 0}^2$ for ordinary self-gravity systems (e.g., Maxwell--Boltzmann distribution). In the limit of  $t \gg T$, 
where $T \equiv 2 \pi / \Omega$ is the oscillation period, it is nicely approximated by
\begin{equation}
\frac{\sin{[(n \Omega - k v_{\rm 0})t]}}{n \Omega - k v_{\rm 0}} \approx \frac{\pi}{k} \delta(v_{\rm 0} - n \Omega / k) \label{dirac_delta} .
\end{equation}
By substituting Equation (\ref{dirac_delta}) with Equation (\ref{partial_integration}), the energy transfer rate per volume is finally 
given by
\begin{equation}
\frac{d K}{dt} = - 2 \pi^2 \sum_{n=1}^{\infty} \frac{A_n^2 \rho_{\rm 0}}{k} \biggl ( \frac{n \Omega}{k} \biggr )^3 \frac{d f(v_{\rm 0})}{d v_{\rm 0}} \biggr |_{v_{\rm 0} = n \Omega / k}. \label{transfer_rate0}
\end{equation}

In actual self-gravitational systems, the phase space density of lower velocity particles is usually dominant, as compared 
with that of higher velocity particles. Because the velocity of resonant particles, $v_{\rm 0} = n \Omega / k$, is positive, 
$d f(v_{\rm 0})/d v_{\rm 0} |_{v_{\rm 0} = n \Omega / k}$ is negative. Thus, $dK/dt$ will be positive. That is, the system will gain 
substantial net energy from the external force.

On the basis of these results, we conclude that the particles satisfying the resonance condition
\begin{equation}
k v_{\rm 0} \approx n \Omega \label{resonance_condition_1}
\end{equation}
effectively gain kinetic energy, and that the system will expand to settle the new equilibrium state, resulting in a decrease 
in density.


\subsection{Resonance Condition}
We apply the resonance condition to spherically symmetric systems. 
The derived formula can be compared with observational results easily and some comparisons are given in Section 4.2. 

First, the wave number as a function of distance, $r$, from the center of the system is redefined by
\begin{equation}
k(r) \equiv \frac{2 \pi}{r}. \label{wavenumber}
\end{equation}
Second, we substitute  $v_{\rm 0}$ in Equation (\ref{resonance_condition_1}) by the typical velocity of particles at $r$ defined by 
\begin{equation}
\sigma(r) = \sqrt{\frac{G M(r)}{r}}, 
\end{equation}
where $M(r)$ is mass within $r$ and $G$ is the gravitational constant. 
Here, we attempt to expand the notion of the critical resonance condition given by Equation (\ref{resonance_condition_1}) as
\begin{equation}
k(r) \sigma(r) \approx n \Omega. \label{resonance_condition_2}
\end{equation}
Then, we define the local dynamical time of the system as
\begin{equation}
t_{\rm d}(r) = \sqrt{\frac{3 \pi}{32 G \bar{\rho}(r)}}, \label{dynamical_time}
\end{equation}
where
\begin{equation}
\bar{\rho} (r) = \frac{M(r)}{({\rm 4} \pi / {\rm 3}) r^3}, \label{rho_bar}
\end{equation}
is the averaged density interior to radius, $r$.  With these applications, the resonant condition becomes
\begin{equation}
t_{\rm d}(r_n) = \frac{\pi}{2\sqrt{2}} \frac{T}{n} \sim \frac{T}{n}, \label{resonance_condition_3}
\end{equation}
where $T$ corresponds the oscillation period for the fundamental tone ($n = 1$) 
and $r_n$ is the radius around which the resonance condition of the $n$th overtone mode is satisfied. 
Considering the Maxwell--Boltzmann DF for particle velocity, Equation (\ref{transfer_rate0}) indicates 
$dK/dt \propto n^4 \exp{(-n^2)}$. The contribution of the fundamental tone to the system will be greater than that of 
the overtone modes. Therefore, we focus on the resonance of the fundamental tone.

The resonance of the overtone modes occurs in the dense region of a DM halo (i.e., around the center), while that of 
the fundamental tone appears in the less-dense region (i.e., outskirts), whose local dynamical time is longer than that 
of the dense region. Thus, the effects of resonances will appear at more inner locations than $r_1$, where the resonance 
condition for the fundamental tone,
\begin{equation}
t_{\rm d}(r_{\rm 1}) \sim T, \label{core_scale}
\end{equation} 
is satisfied. 
By using the inversion procedure to solve this equation, the resonance radius is obtained by
\begin{equation}
r_1=t_d^{-1}(T). 
\label{core_scale_inv}
\end{equation} 

In the following subsection, we apply this prescription of the resonant cusp--core transition to the models of CDM halos 
predicted by cosmological $N$-body simulations, and estimate the core scale.


\subsection{Core Scale}
The mass density profile of a DM halo obtained from cosmological $N$-body simulations based on the CDM scenario can be fitted by the following formula:
\begin{equation}
\rho_{\rm DM}(x) = \frac{\rho_0}{x^{\alpha}(x+1)^{3-\alpha}}, \label{DM_halo}
\end{equation}
where $x \equiv r/R_{\rm DM}$, $r$ is the distance from the center of a DM halo; $\alpha$ is the power--law index of 
the central cusp; and  $\rho_0$ and $R_{\rm DM}$ are the scale density and scale length of a DM halo, respectively. 
This formula indicates that the density distribution of a DM halo changes from $\rho \propto r^{-\alpha}$ in the center
($r < R_{\rm DM}$) to $\rho \propto r^{-3}$ at the outskirts ($r > R_{\rm DM}$). Here, $\alpha =1.0$ corresponds to 
the Navarro--Frenk--White (NFW) model (Navarro et al. 1996b; Navarro et al. 1997), and $\alpha =1.5$ corresponds to 
the Fukushige--Makino--Moore (FMM) model (Fukushige \& Makino 1997; Moore et al. 1999).

The mass profile derived by the volume integration of Equation (\ref{DM_halo}) is given by
\begin{equation}
M(\alpha; x) = \frac{4 \pi \rho_0 R_{\rm DM}^3}{3 - \alpha} x^{3 - \alpha} \,_2F_1 [3-\alpha, 3-\alpha, 4-\alpha; -x], \label{mass_profile}
\end{equation}
where $_{\rm 2}F_{\rm 1} [3-\alpha, 3-\alpha, 4-\alpha; -x]$ is Gauss's hypergeometric function, which is hereafter denoted as
$_{\rm 2}F_{\rm 1}$ [$\alpha; -x$]. From Equation (\ref{mass_profile}), 
\begin{equation}
\rho_0 = \frac{3-\alpha}{4 \pi R_{\rm DM}^3} \frac{1}{c^{3-\alpha} {_{\rm 2}F_{\rm 1}} [\alpha; -c]} M_{\rm vir}. \label{scale_density}
\end{equation}
The virial mass of a DM halo is defined by
\begin{equation}
M_{\rm vir} = \frac{4 \pi}{3} \rho_{\rm cri} (1+z)^3 \Delta R_{\rm vir}^3,
\label{virial_mass}
\end{equation}
where $\rho_{\rm cri}$ is the critical density of the universe, $z$ is the redshift, $\Delta(=200)$ is a parameter of density 
enhancement, $c=R_{\rm vir}/R_{\rm DM}$ is a concentration parameter, and $R_{\rm vir}$ is the virial radius.

By using Equations (\ref{mass_profile}) and (\ref{scale_density}), the averaged density of a DM halo interior to radius $x$ is given by
\begin{equation}
\bar{\rho} (x) = \frac{3 M_{\rm vir}}{4 \pi R_{\rm DM}^3} \frac{x^{-\alpha}}{c^{3-\alpha}} \frac{_{\rm 2}F_{\rm 1} [\alpha; -x]}{  _{\rm 2}F_{\rm 1} [\alpha; -c]},
\end{equation}
and by applying the resonance condition given in Equation (\ref{core_scale}), we finally arrive at the core scale 
$r_{\rm core}=x_{\rm core} \, R_{\rm DM}$. 

We now discuss the utmost limit, $x_{\rm core} \ll {\rm 1}$.
In this case, the approximation $_{\rm 2}F_{\rm 1} [\alpha; -x] \sim {\rm 1}$ holds. 
Thus, the core scale is expressed as
\begin{eqnarray}
r_{\rm core} &=& R_{\rm DM} \biggl [ \frac{8 G}{\pi^2} \frac{M_{\rm vir} T^2}{R_{\rm DM}^3 c^{3 - \alpha} {_{\rm 2}F_{\rm 1} [\alpha; -c]}} \biggr ]^{1/\alpha}, \\
          &=& R_{\rm DM} \biggl [ \frac{32 G}{3\pi} \Delta \rho_{\rm cri} (1+z)^3 \frac{c^\alpha}{_{\rm 2}F_{\rm 1} [\alpha; -c]} T^2 \biggr ]^{1/\alpha}.
\label{2f1_core_scale}
\end{eqnarray}

%
%
%
%

\begin{table}
\begin{center}
\caption{Numerical Values of ${_{\rm 2}F_{\rm 1} [\alpha; -c]}$ for Conventional Models of DM Halos}
\begin{tabular}{c|cccc}
$\alpha$ (Model)& $c = 5$ & $c = 10$ & $c = 15$ & $c = 20$ \\
\tableline\tableline
1.0 (NFW)       & 0.07667 & 0.02978  & 0.01631  & 0.01046  \\ 
1.5 (FMM)       & 0.16948 & 0.08681  & 0.05656  & 0.04117  \\ 
\tableline
\end{tabular}
\tablecomments{$\alpha$ is the power--law index of the central cusp of the DM halo and $c$ is the concentration parameter.}
\label{2f1_values}
\end{center}
\end{table}

By using Equation (\ref{virial_mass}), more convenient formulae are given by
\begin{eqnarray}
r_{\rm core} = &0.74& \ {_{\rm 2}F_{\rm 1} [1; -c]^{-1}} (1+z)^2 \nonumber \\ 
              &\times& \bigl ( \frac{M_{\rm vir}}{10^9 M_{\odot}}\bigr )^{1/3} \bigl ( \frac{T}{10^7 {\rm yr}} \bigr )^2 {\rm pc}, \label{nfw_general} 
\end{eqnarray}
for the NFW model, and 
\begin{eqnarray}
r_{\rm core} = &23& \ {_{\rm 2}F_{\rm 1} [1.5; -c]^{-2/3}} (1+z) \nonumber \\ 
              &\times& \bigl ( \frac{M_{\rm vir}}{10^9 M_{\odot}}\bigr )^{1/3} \bigl ( \frac{T}{10^7 {\rm yr}} \bigr )^{4/3} {\rm pc}, \label{fmm_general}
\end{eqnarray}
for the FMM model.  Here, we show some values of ${_{\rm 2}F_{\rm 1} [\alpha; -c]}$ in Table \ref{2f1_values} for 
$\alpha = {\rm 1.0, 1.5}$ and some ordinary $c$ parameters obtained from cosmological $N$-body simulations (e.g., 
Macci{\`o} et al. 2008). Moreover, assuming $c={\rm 10}$ and $z={\rm 0}$, we reach the core scale represented by the following 
simpler formulae:
\begin{eqnarray}
r_{\rm core} = 25 \ \bigl ( \frac{M_{\rm vir}}{10^9 M_{\odot}}\bigr )^{1/3} \bigl ( \frac{T}{10^7 {\rm yr}} \bigr )^2 {\rm pc}, \label{nfw_simple}
\end{eqnarray}
for the NFW model, and 
\begin{eqnarray}
r_{\rm core} = 120 \ \bigl ( \frac{M_{\rm vir}}{10^9 M_{\odot}}\bigr )^{1/3} \bigl ( \frac{T}{10^7 {\rm yr}} \bigr )^{4/3} {\rm pc},  \label{fmm_simple}
\end{eqnarray}
for the FMM model.

In the following section, we evaluate the quality and integrity of the analytical prediction by performing numerical 
experiments of $N$-body simulations.

%
%
%
%

\section{NUMERICAL SIMULATIONS}


\subsection{Numerical Models}
We develop a parallelized tree code for graphics processing unit (GPU) clusters that adopts the standard algorithm proposed 
by Barnes \& Hut (1986) and employs the second-order Runge--Kutta scheme as the time integration method. In light of 
Nakasato (2012), we design CPU cores to compute tree construction and GPU cards to compute tree traversal 
(Ogiya et al. 2013). To generate $N$-body systems with a cusp described in Equation (\ref{DM_halo}), in the equilibrium state, 
we use the fitting formulation of the DF proposed by Widrow (2000). In this case, the DF depends only on energy and 
the velocity dispersion of the system is isotropic.

We simulate the dynamical response of a DM halo with virial mass, $M_{\rm vir} = 10^9 M_{\odot}$, virial radius, $R_{\rm vir} = 10$,
and scale length, $R_{\rm DM} = 2$ kpc, assuming the formation redshift, $z =1$. Here, we define $\tau=10$ Myr for the NFW model 
and 4 Myr for the FMM model as the typical timescales of dynamical response near the innermost regions of DM halos. 
These values correspond to $t_{\rm d}({\rm 50 pc})$ for respective models. 
Throughout this paper, the tolerance parameter of the Barnes--Hut tree algorithm (Barnes \& Hut 1986) is $\theta = 0.6$ 
and the softening parameter is $\epsilon = 0.004$ kpc. The total number of particles is $N = 16, 777, 216$ for ordinary runs. 
In the high-resolution run for the check of numerical convergence, we use $N = 134, 217, 728$.

%
%
%
%

\begin{figure}
\epsscale{1.15}
\plotone{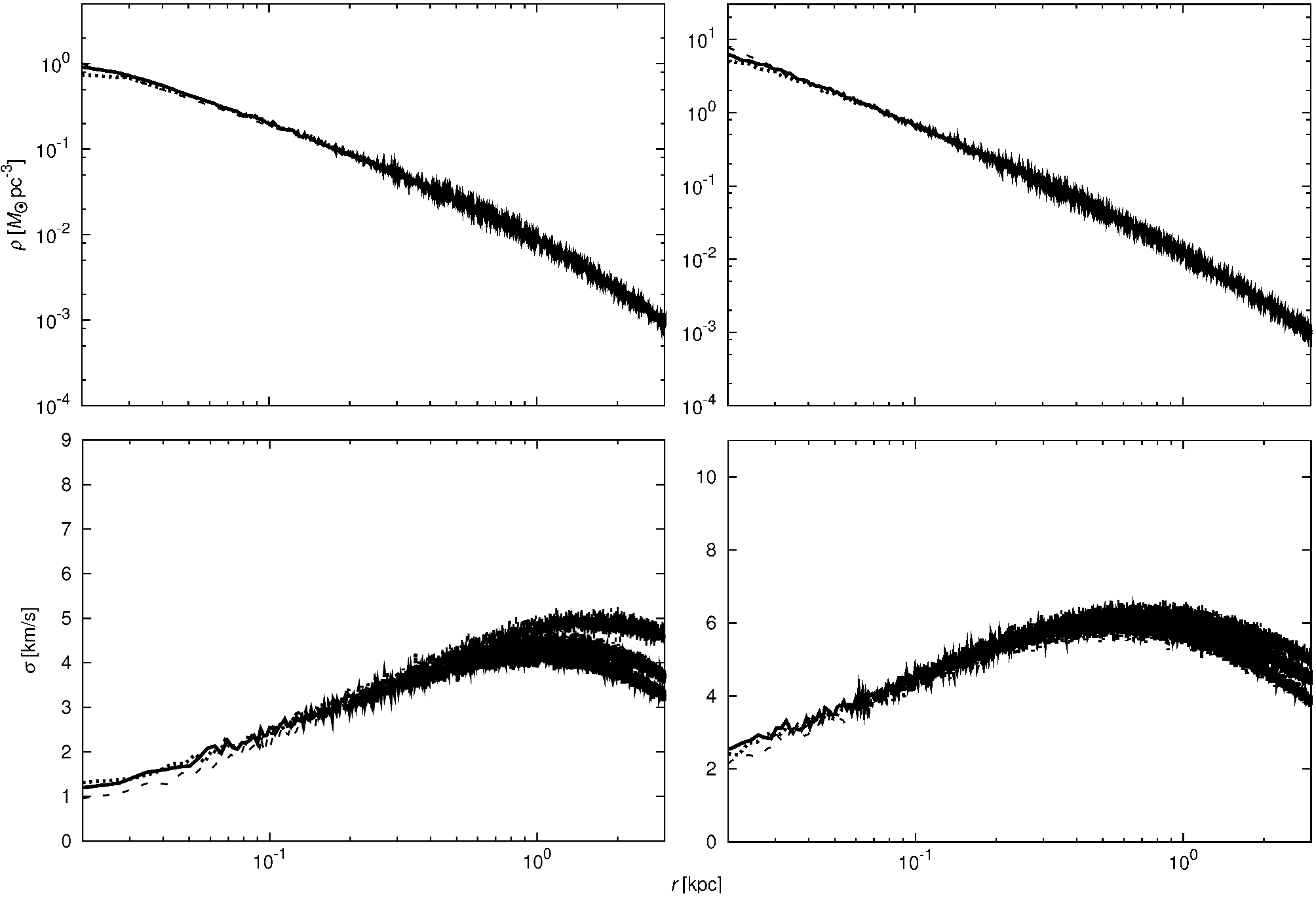}
\caption{
Density and velocity dispersion profiles of nonperturbed (equilibrium) NFW (left) and FMM (right) halos. Dashed lines in 
the upper panels (density profile) are the analytical formulation given by Equation (\ref{DM_halo}). In the lower panels (velocity 
dispersion), these lines represent the DM halos created by following the formula described in the text. 
Even after 100 $\tau$ (solid lines) and 200 $\tau$ (dotted lines), our $N$-body models remain the initial configurations.
\label{non_perturb}
}
\end{figure}

The baryon potential near the centers of DM halos is represented by the Hernquist-type potential (Hernquist 1990), which is given by
\begin{equation}
\Phi_{\rm b} (r, t) = -\frac{G M_{\rm b}}{r+R_{\rm b}(t)}, \label{Hernquist}
\end{equation}
where $M_{\rm b}$ and $R_{\rm b}(t)$ are the total baryon mass and the scale length of the external potential, respectively. 
We change the scale length of the baryon potential to 
$R_{\rm b}(t) = {\rm 0.5} [R_{\rm b, max} - R_{\rm b, min}] [{\rm 1} + \cos{(\Omega t - \phi_{\rm 0})}] + R_{\rm b, min}$, 
$T$ is the timescale of oscillation of the external potential,  $R_{\rm b, max}$ and $R_{\rm b, min}$ are the maximum and minimum 
scale lengths, and $\phi_{\rm 0}$ is the initial phase of oscillation, respectively. Note that even if no external 
potential exists, the cusp-to-core transition may occur by two-body relaxation for a few particles to resolve the central 
cusp. Figure \ref{non_perturb} clearly shows that in our model, the system remains stable for the entire time and that the effect 
of two-body relaxation is not observed for at least several $100\tau$.

To construct a stable initial condition, we dynamically relax the equilibrium $N$-body system in the abovementioned external 
baryon potential with the fixed scale length $R_{\rm b} = R_{\rm b, 0}$ for $\sim 30 \tau$ where $R_{\rm b, 0}$ is the initial 
scale length of the baryon potential. We define the initial phase of oscillation as 
$\phi_{\rm 0} = \arccos{[2(R_{\rm b, 0} - R_{\rm b, min})/(R_{\rm b, max} - R_{\rm b, min})-1]}$.

%
%
%
%

\begin{table*}
\begin{center}
\caption{Summary of Simulation Runs}
\begin{tabular}{ccccccc}
ID & $T $ [ $\tau$ ] & $M_{\rm b}$ [ $\times 10^8 M_{\odot}$ ] & $R_{\rm b, 0}$ [kpc] & $R_{\rm b, max}$ [kpc] &  $R_{\rm b, min}$ [kpc] \\
\tableline\tableline
NFW--NP     & --       & --   & --  & --  & --   \\  
NFW--1       & 1        & 1.7  & 1.0 & 2.0 & 0.04 \\ 
NFW--2       & 2        & 1.7  & 1.0 & 2.0 & 0.04 \\ 
NFW--3, 3HR & 3        & 1.7  & 1.0 & 2.0 & 0.04 \\ 
NFW--4       & 4        & 1.7  & 1.0 & 2.0 & 0.04 \\
NFW--5       & 5        & 1.7  & 1.0 & 2.0 & 0.04 \\ 
NFW--6       & 10       & 1.7  & 1.0 & 2.0 & 0.04 \\ 
NFW--7       & 1 \& 10  & 1.7  & 1.0 & 2.0 & 0.04 \\
NFW--8       & 1        & 0.85 & 1.0 & 2.0 & 0.04 \\
NFW--9       & 1        & 1.7  & 1.0 & 2.0 & 0.2  \\
NFW--10     & 1        & 1.7  & 0.1 & 0.2 & 0.04 \\
NFW--11     & 1        & 1.7  & 0.1 & 2.0 & 0.04 \\
FMM--NP    & --       & --   & --  & --  & --   \\
FMM--1      & 1        & 1.7  & 1.0 & 2.0 & 0.04 \\ 
FMM--2      & 2        & 1.7  & 1.0 & 2.0 & 0.04 \\ 
FMM--3      & 3        & 1.7  & 1.0 & 2.0 & 0.04 \\ 
FMM--4      & 4        & 1.7  & 1.0 & 2.0 & 0.04 \\ 
FMM--5      & 5        & 1.7  & 1.0 & 2.0 & 0.04 \\ 
FMM--6      & 10       & 1.7  & 1.0 & 2.0 & 0.04 \\ 
\tableline
\end{tabular}
\tablecomments{
NFW and FMM indicate the initial DM halo model. NP is the nonperturbed run for testing the stability of the $N$-body systems. 
$T$ and $M_{\rm b}$ are the oscillation period and total mass of the baryon potential, respectively. $R_{\rm b, 0}$, $R_{\rm b, max}$, and $R_{\rm b, min}$ 
are initial, maximum, and minimum scale length of the external baryon potential, respectively. We adopt the mass ratio of 
the baryon component to the DM halo cosmic value obtained from $Wilkinson \ Microwave \ Anisotropy \ Probe$ observations 
(Spergel et al. 2007; Komatsu et al. 2009, 2011). 
}
\label{parameters}
\end{center}
\end{table*}

%
%
%
%

\begin{figure}
\epsscale{1.15}
\plotone{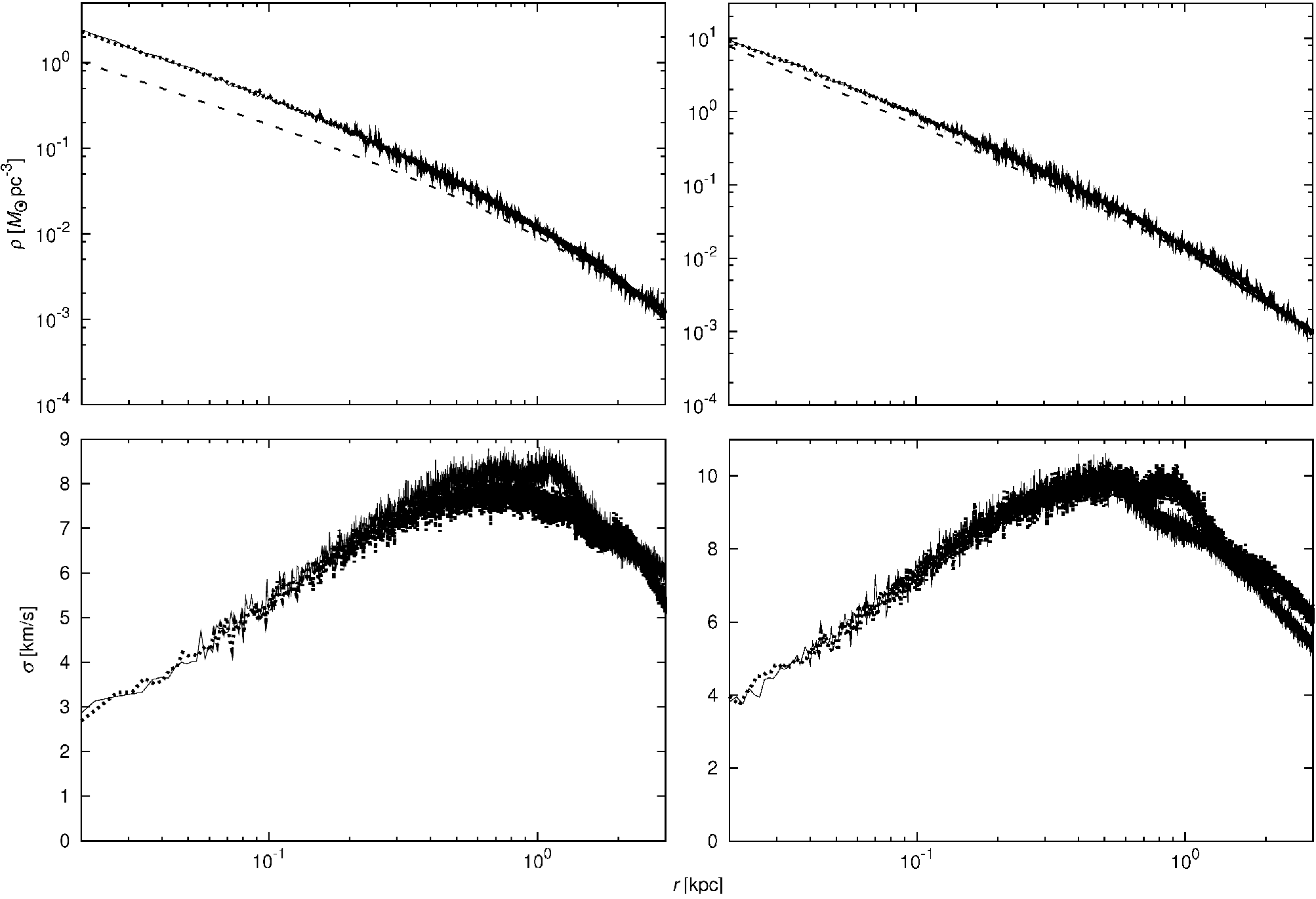}
\caption{
State of the dynamically relaxed NFW (left) and FMM halo (right) with the static external potential for $R_{\rm b}=R_{\rm b,0}$. 
Top and bottom panels show the density and velocity dispersion profiles of DM halos, respectively. The thin solid line 
(thick dotted line) represents $t =15\tau (t = 30\tau)$ after adding the static external potential for a constant scale length 
$R_{\rm b}(t)=R_{\rm b, 0}=1$ kpc defined by Equation (\ref{Hernquist}). 
\label{with_fixed_potential}
}
\end{figure}

The left (right) panel of Figure \ref{with_fixed_potential} indicates the states of dynamically relaxed NFW (FMM) halos. 
The top and bottom panels represent the density and velocity dispersion profiles of the DM halos, respectively. The thin solid 
line corresponds to the snapshot of the state exposed to the static external potential for $15 \tau$. After $30 \tau$, 
the system reaches the quasistable state and remains virtually unchanged (dotted line). Because of the static external potential, 
the DM halos show a slight contraction (top panels). A comparison between the bottom panels of Figures \ref{non_perturb} and 
\ref{with_fixed_potential}  reveals that the velocity dispersion becomes larger than the equilibrium state because of 
the deeper potential. 
Table \ref{parameters} provides a summary of our simulation runs.


\subsection{Results}
In this section, we present the results of our $N$-body simulations and compare them with the theoretical predictions of 
the simple analytic model given in Section 2.

%
%
%
%

\begin{figure}
\epsscale{1.15}
\plotone{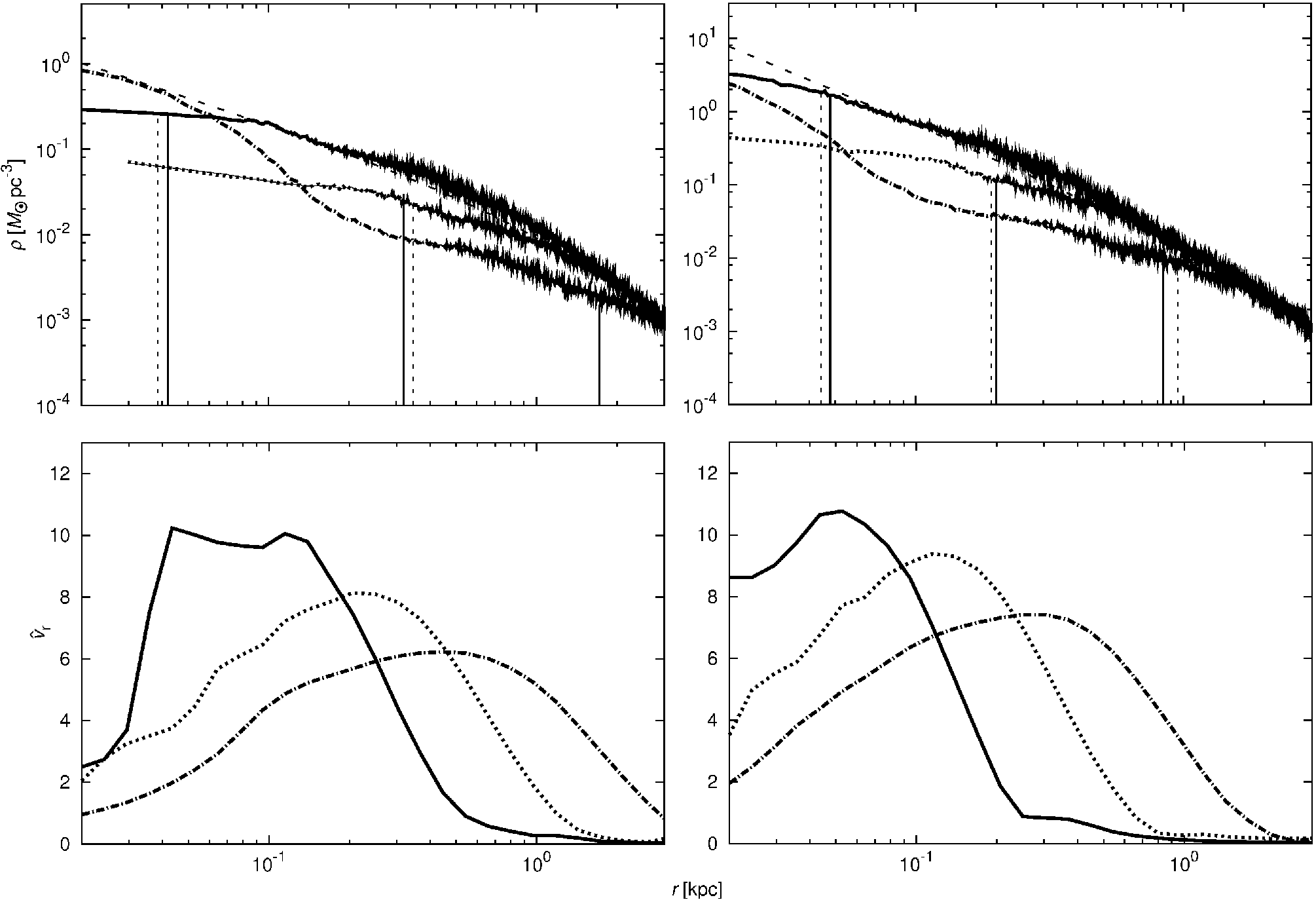}
\caption{
Density profile (top panels) and spectrum of radial velocity (bottom panels) of DM halos after 10 oscillations of the external 
potential. Left panels show results of the NFW model (NFW--1, NFW--3, and NFW--6), and right panels show those of the FMM model 
(FMM--1, FMM--3, and FMM--6). Solid, dotted, dotted--dashed, and dashed lines represent $T=\tau, 3\tau$, and $10\tau$ and 
the initial condition, respectively. The thin solid line in the upper left panel represents the high-resolution run of 
$T = 3\tau$ (NFW--3HR). Each vertical and solid line indicates the core scale predicted by Equation (\ref{core_scale_inv}). 
The dashed ones are derived using Equations (\ref{nfw_general}) or (\ref{fmm_general}).
\label{T_dependence_spectrum}
}
\end{figure}

Figure \ref{T_dependence_spectrum} demonstrates how the dynamical responses of DM halos depend on the oscillation frequency 
of the external baryon potential. In each panel, solid, dotted, dotted--dashed, and dashed lines correspond to $T = \tau, 3\tau$, 
and $10 \tau$ and the initial condition, respectively.
The thin solid line represents the high-resolution run of $T = 3 \tau$. It is clearly shown that the results converge very well.
The upper left (upper right) panel shows the resultant density profiles of DM halos for the initial NFW (FMM) model after 10 
oscillation cycles. 
Models of $T=\tau$ and $T=3\tau$ clearly show the cusp--core transition, and the resultant core scale depends on 
the oscillation frequency of the external baryon potential. 
In the runs of $T =10\tau$, the bump structures, decreasing in density at the outskirt with a remaining central cusp, appear.
As discussed in Section 2, the overtone modes affect the center region, 
and the impacts of the overtones of high Fourier indexes are weaker than those of the fundamental tone and overtones of low indexes. 
Therefore, the number of oscillation cycles is not sufficient to flatten the cusp for the runs.

%
%
%
%

\begin{figure}
\plotone{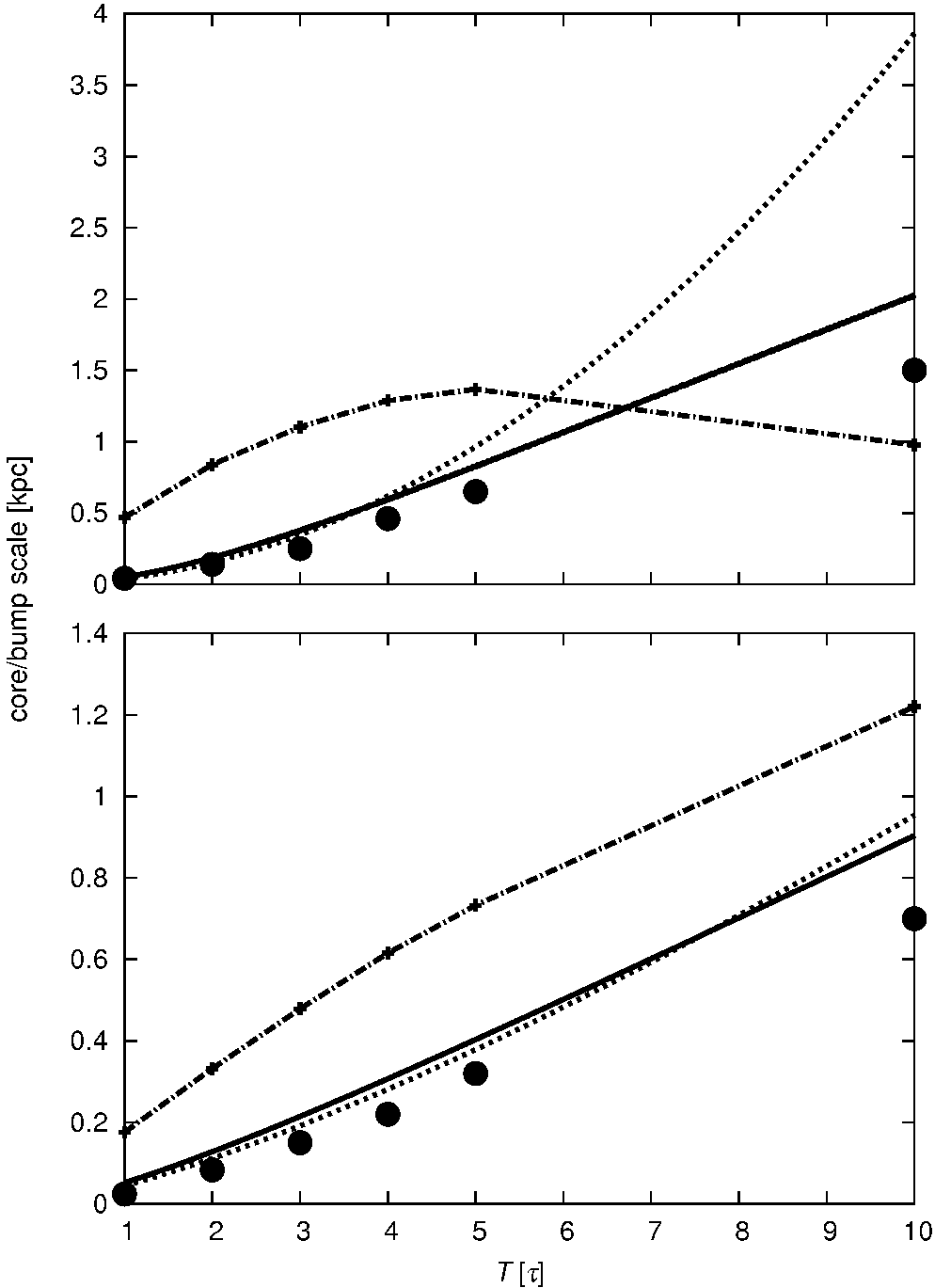}
\caption{
Comparison of resultant core or bump scale in simulations, $r_{\rm sim}$ with analytical predictions. Upper and lower panels show results of NFW and FMM models, respectively. Filled circles represent $r_{\rm sim}$ defined as the outermost point at which the obtained density in simulations is equal to or less than the half of the initial density. Solid, dotted and dotted--dashed lines show predictions by Equation (\ref{core_scale_inv}), $r_{\rm 1}$, by Equation (\ref{nfw_general}) or (\ref{fmm_general}), $r_{\rm core}$, and by Equation (\ref{transfer_rate}) for 10 oscillation cycles, $r_{\rm up}(10)$, respectively. 
\label{comparison}
}
\end{figure}

The vertical dashed lines in both panels represent the core scales predicted by Equations (\ref{nfw_general}) and (\ref{fmm_general}). 
The analytical predictions based on Equation (\ref{core_scale_inv}) are indicated by vertical solid lines. 
Figure \ref{T_dependence_spectrum} shows that our analytic model well 
matches the core scale derived by numerical 
experiments for runs of $T = \tau$ and $3 \tau$. 
The predicted resonance point for runs of $T = 10 \tau$ reasonably matches the density decreasing region.  
Although it is beyond the appropriate range of the approximation $x_{\rm core} \ll 1$, Equation (\ref{fmm_general}) gives a reasonable prediction for the bump structure of the FMM halo. 
Figure \ref{comparison} provides comparisons between the resultant core or bump scales in $N$-body simulations and theoretical predictions. 
In simulations, the core or bump scale is defined as the outermost point at which the obtained density is equal to or less than the half of the initial density. 
As shown by the solid and dotted lines, the resonance model predicts the scale of decreasing density well (i.e., core or bump scales).

We computed the Fourier transformation of the radial velocity to verify the emergence of the energy transfer due to 
the resonance between the particles and the density waves. Because DM halos are perturbed radially in our model, the induced 
density waves will propagate in that direction. The Fourier transform technique is useful for decomposing wave components. 
By using snapshot data of a given time, $t$, we compute the radial velocity profile, $v_{\rm r}(r, t)$, which is the averaged 
radial velocity of particles in each bin. Then, $v_{\rm r}$ is transformed into the temporal Fourier components (i.e., spectrum),
${\hat v}_{\rm r}(r, \omega)$, by the conventional procedure,
\begin{equation}
{\hat v}_{\rm r}(r, \omega) = \int v_{\rm r}(r, t) e^{-i \omega t} dt, \label{fourier_transformation}
\end{equation}
with the sampling frequency of $0.1\tau$. The lower panels in Figure \ref{T_dependence_spectrum} show that ${\hat v}_{\rm r}$ 
of the angular frequency, $\omega$, equals the oscillation frequency of the external potential $\Omega \equiv 2 \pi / T$. 
The positions of the peaks clearly match the core scale in the density profile and depend on $T$. The figure demonstrates 
the spectrum of the fundamental mode for each run. These results clearly show that, as expected, the core scale agrees with 
the peak's position predicted by Equation (\ref{core_scale_inv}). As discussed in the previous section, the resonance of the slow 
oscillation appears at the less-dense region of the DM halo (i.e., outskirts), whereas that of the rapid oscillation appears 
at the central denser region. This result indicates that when DM halos are in resonance with density waves induced by the external potential, particles 
are accelerated effectively so that the 
density is decreased near the peak.

%
%
%
%

\begin{figure}
\epsscale{1.15}
\plotone{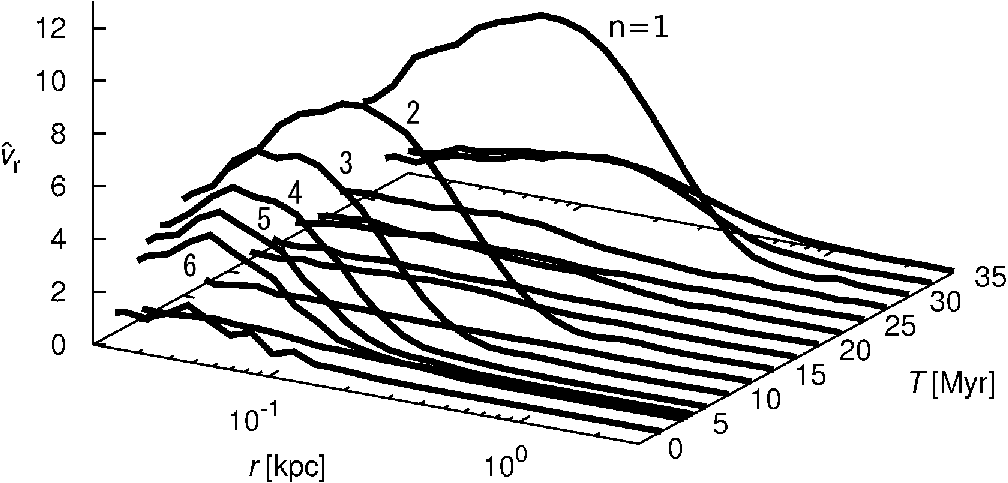}
\caption{
Spectrum of radial velocity ${\hat v}_{\rm r}$ of the DM halo (NFW--3) including the case of  $\omega \neq \Omega$. 
Each line represents ${\hat v}_{\rm r}$ for different frequencies, and $\omega$ is shown as the period, $T$ 
(viz. $\omega = 2 \pi / T$). The indices of the overtone modes, $n$, are denoted by the corresponding lines.
\label{spectrum}
}
\end{figure}

Figure \ref{spectrum} shows ${\hat v}_{\rm r}$, including the case of $\omega \neq \Omega$. Although only the result of NFW--3 
is shown in this figure, a similar phenomenon appears for other models. In the bottom panels of Figure \ref{T_dependence_spectrum}, 
we show only the fundamental mode for each $T$. Figure \ref{spectrum} explains the resonances of the overtone modes expected 
from Equation (\ref{resonance_condition_1}); these appear with peaks, and their overtone indexes $n$ are denoted. These peaks never 
appear outside the modes with integral indexes. The figure also shows a tendency that the signatures of the resonances of higher 
overtones are fainter than those of lower overtones or the fundamental tone, which implies that the influences of the resonances 
of the overtones weaken with an increase in $n$, as expected in Section 2. In the runs of $T =10\tau$, the 10th overtone may contribute 
to particle acceleration; however, its impact is not sufficient to flatten the central cusp. On the contrary, in the runs of 
$T = 3\tau$, the effects of the third overtone are significant to the systems, and the core structure is created at the center. 
Because its fundamental tone is effective near $r \sim $ 340 and 190 pc for NFW and FMM models, respectively, the core scale 
becomes larger than that of $T =\tau$.

%
%
%
%

\begin{figure}[h]
\plotone{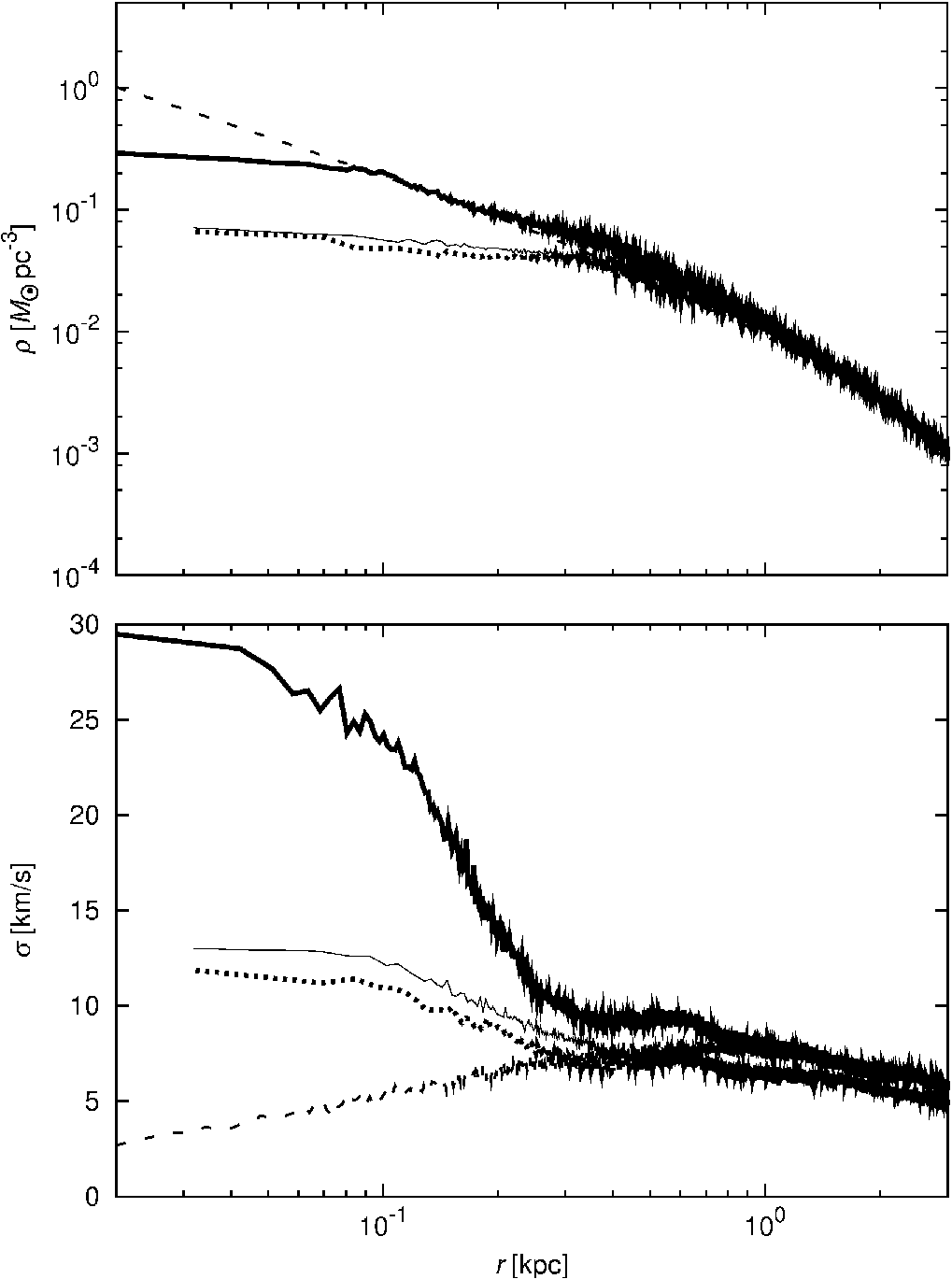}
\caption{
Density and velocity dispersion profiles of the DM halo (NFW--1) after oscillation has stopped. The thick solid line represents 
the state at the end of 10 oscillation cycles. We then add the static external potential for $R_{\rm b}=R_{\rm b,0}=1$ kpc to 
the system. The dotted and thin solid lines express the snapshots after $10\tau$ and $50\tau$. 
Dashed line in the top panel is the NFW profile. 
That in the bottom panel represents the dynamically relaxed NFW halo with the static external potential for $R_{\rm b}=1$ kpc. 
\label{stop_oscillation}
}
\end{figure}

We next determine whether the central cusp recovers when the gas oscillation stops. To review the dynamical evolution of 
the DM halo in such cases, we extend the simulation of NFW--1 from the snapshot after 10 oscillation cycles. In this extension, 
we add the static external potential for $R_{\rm b}=R_{\rm b,0}=1$ kpc, which corresponds to the gravitational potential of 
the quiescent baryon component. Figure  \ref{stop_oscillation} clearly shows that after the gas oscillation has stopped, 
the central cusp does not recover, despite the attracting force of the static external potential. Because of the heating 
by the oscillatory potential, the system expands and the core scale increases after the oscillation process ends. 
The DM halo then reaches the new quasiequilibrium state.

The cusp remains when the oscillation period is sufficiently longer than the local dynamical time near the center (dotted--dashed 
line in Figure \ref{T_dependence_spectrum}). Although we have thus far assumed mono-periods for simplicity, multiperiodic 
oscillations occur in real galaxies. Because the resonance scale is determined by the period, such oscillations are expected 
to show effects in multiscales, and the remaining cusp may be erased. To test this hypothesis, we conduct an $N$-body 
simulation (NFW--7). In this run, we modify the external potential as
\begin{equation}
\Phi_{\rm b} (r, t) = -\frac{G M_{\rm b}}{2} \biggl[ \frac{1}{r+R_{\rm b,1}(t)} + \frac{1}{r+R_{\rm b,2}(t)} \biggr] .
\label{mul_freq_pot}
\end{equation}
Here, $R_{\rm b,1}(t)$ and $R_{\rm b,2}(t)$ are the scale length of the external potential and have periods 
$T_1 = \tau = {\rm 10Myr}$ and $T_2 = 10 \tau = {\rm 100Myr}$, respectively.

%
%
%
%

\begin{figure}
\plotone{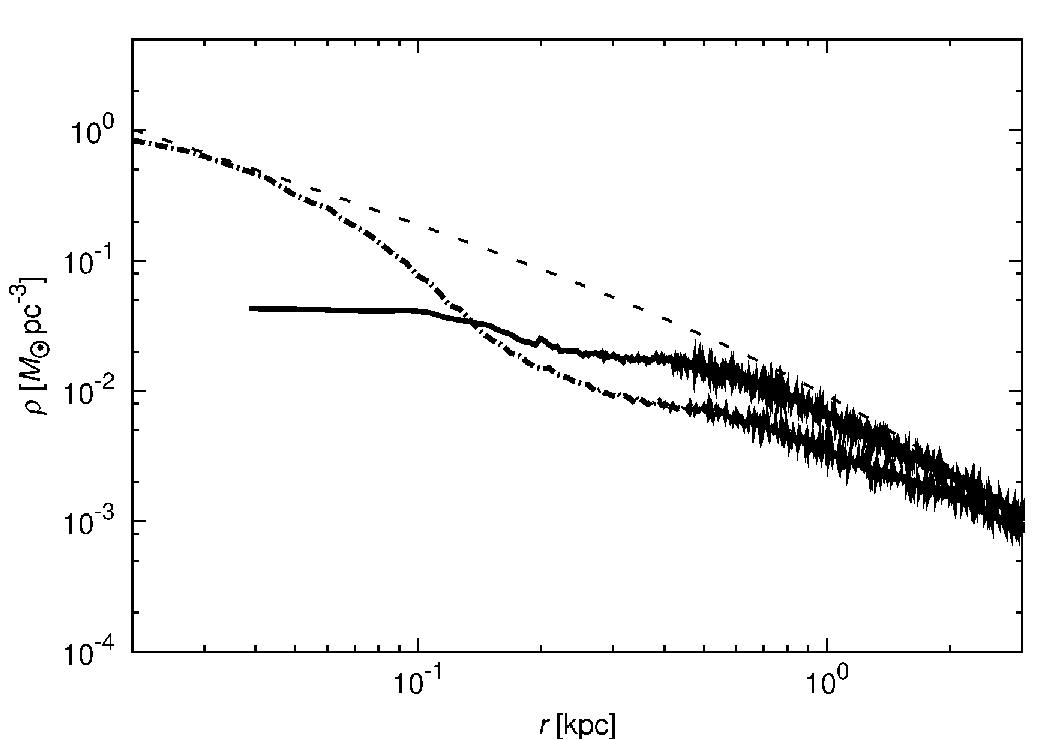}
\caption{
Density profile of the DM halos after exposure to the external potential for 1 Gyr. Solid and dotted--dashed lines represent 
the results of NFW--7 and NFW--6, respectively. The dashed line represents the NFW profile.
\label{multi_freq} 
}
\end{figure}

In Figure \ref{multi_freq}, we compare the results of this simulation with those of NFW--6. 
The central cusp is flattened in 
the run of NFW--7 (solid line). Interestingly, the core size is almost the same as the bump scale of NFW--6. As discussed in 
Section 2, the core size is determined by the resonance of the slowest mode. In the runs of NFW--6 and NFW--7, the slowest modes are 
the same ($T = 100$ Myr). Conversely, the density structures at the innermost regions differ significantly because of 
the strength of the rapid mode resonation near the center, which may have flattened the cusp. As described in Section 2, 
although the external potential in NFW--6 contains the modes' affect at the center, they are higher overtone modes and 
their strengths decrease with the index of Fourier components, $n$. In NFW--7, however, the mode of $T = 10$ Myr is included 
as one of the fundamental tones, and its strength is sufficiently large to significantly affect the DM halo. 
Consequently, the resonance of the rapid mode, $T_1 = \tau$, destroys the cusp, and that of the slow mode, $T_2 = 10 \tau$, 
expands the core size in this run.

%
%
%
%
\section{Discussion}

\subsection{Energy Transfer Rate and Number of Cycles}
We have demonstrated that the density profile of DM halos is significantly affected by the resonance between the DM particles 
and the density waves excited by the oscillation of the interstellar gas. Although the oscillation period is the only focus of the previous section, 
the mass of gas component  ($M_{\rm b}$) and width of gas oscillation ($R_{\rm b,min}, R_{\rm b,max}$) are also important factors 
in the real galaxies because they determine the amplitude of potential change and the number of oscillation cycles required 
to flatten the cusp. Here, we discuss such dependence by using the argument of energy transfer rate from the oscillatory change 
in galactic potential to DM halos.

We modify the energy transfer rate given by Equation (\ref{transfer_rate0}) as
\begin{equation}
\frac{d K(r)}{dt} = - 2 \pi^2 \sum_{n=1}^{\infty} \frac{A_n(r)^2 \bar{\rho}(r)}{k(r)} v_{\rm 0}^3(r) \frac{d f(v_{\rm 0})}{d v_{\rm 0}}(r) \biggr |_{v_{\rm 0}(r) = n \Omega / k(r)}, \label{transfer_rate}
\end{equation}
where  $v_{\rm 0}(r) = n \Omega / k(r)$ is the velocity of particles resonating with the $n$th overtone mode. The DF of 
the particle velocity, $f(v)$, is adopted by the fitting function proposed by Widrow (2000), and  $dK(r)/dt$ is the kinetic 
energy density injected from the external force to the system per unit time. 
The estimated energy transfer rate should be larger than the results of three-dimensional simulations because we assume that the kinetic energy is transfered from baryon to the DM halo with a constant rate, which is estimated from the physical values of the initial equalibrium state. The energy transfer rate is expected to damp with time since the resonating particles are accelerated and the number of such particles decreases through the resonance. 
The core size estimated by the energy transfer rate corresponds to the upper limit of that created through resonances.  
We compare these values with the binding energy density of the equilibrium NFW halo, $W(r)$, to estimate the number of oscillation cycles required for excess injected kinetic energy over the binding energy, $N_{\rm osc}$. 
Dotted--dashed lines in Figure \ref{comparison} provide a comparison between simulation results and predictions based on this analysis and show that the predictions tend to be larger than simulation results as discussed above.

%
%
%
%

\begin{figure}
\plotone{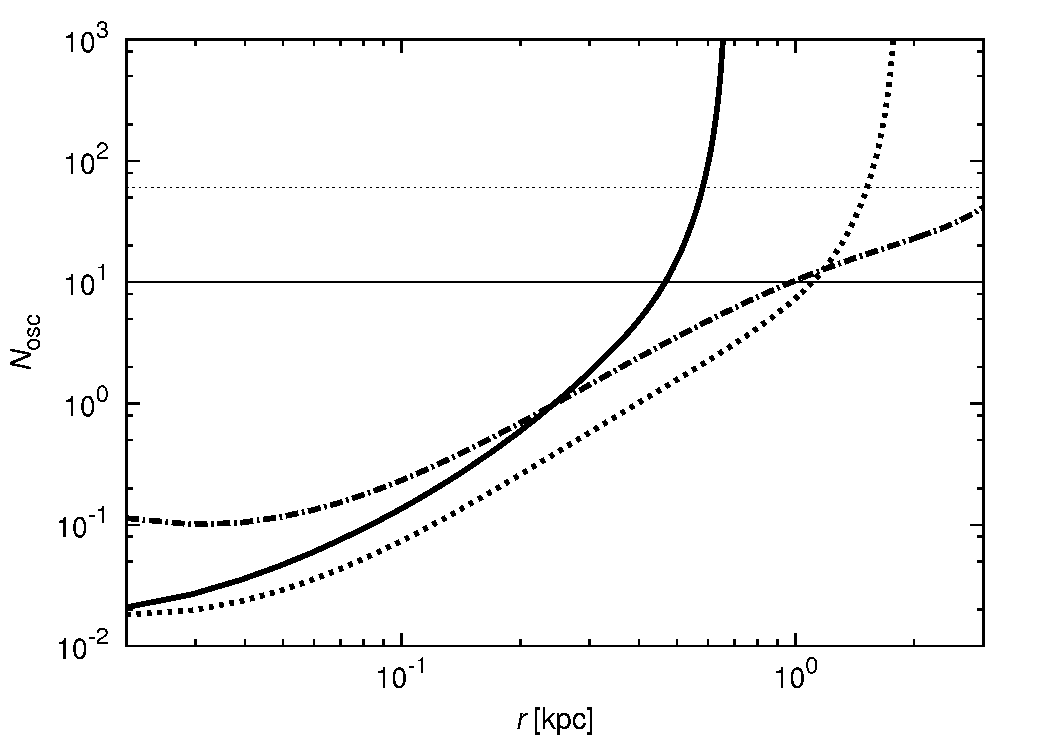}
\caption{
Number of oscillation cycles for the excess of injected kinetic energy over the binding energy, $N_{\rm osc}$, as a function 
of $r$. Thick solid, dotted, and dotted--dashed lines represent  $T = \tau, 3\tau$, and $10\,\tau$, respectively. 
Horizontal solid (dotted) lines represent 10 (60) cycles.
\label{etr_standard}
}
\end{figure}

Figure \ref{etr_standard} shows the core scale predicted by this model. In these runs, we set the number of oscillation 
cycles at 10. According to this estimation, the inputted kinetic energy will be exceeded in the region of curves below 
the horizontal solid line, and the core structure will be created there. The results of the simulation runs for $T = \tau$ 
and $3\tau$ (NFW--1, 3) accommodate the prediction (upper left panel in Figure \ref{T_dependence_spectrum}). The estimation for 
the core scale, determined using the energy transfer rate, is consistent with the prediction based on the resonant condition 
derived in Section 2 because it corresponds to the position at which particles resonate with the fundamental tone and are accelerated most effectively. 
However, our model, based on the energy transfer rate, tends to predict a slightly larger core scale than that shown in the simulation results, which may have been caused by the overestimation of the rate of kinetic energy transfer from the baryon potential to the DM 
halo because of the approximation from Equation (\ref{dirac_delta}). Nonlinear or multidimensional effects not considered in 
the model may also have led to the overestimation. Our model fails the prediction for the run of $T =10\tau$, which is 
unfortunate because the interactions among different Fourier modes that we neglect are essential in this case. Particles moving 
faster than density waves push them and lose kinetic energy. Evaluation of such power to accurately estimate the net energy 
transfer rate requires a significantly more complex analysis.

%
%
%
%

\begin{figure}
\plotone{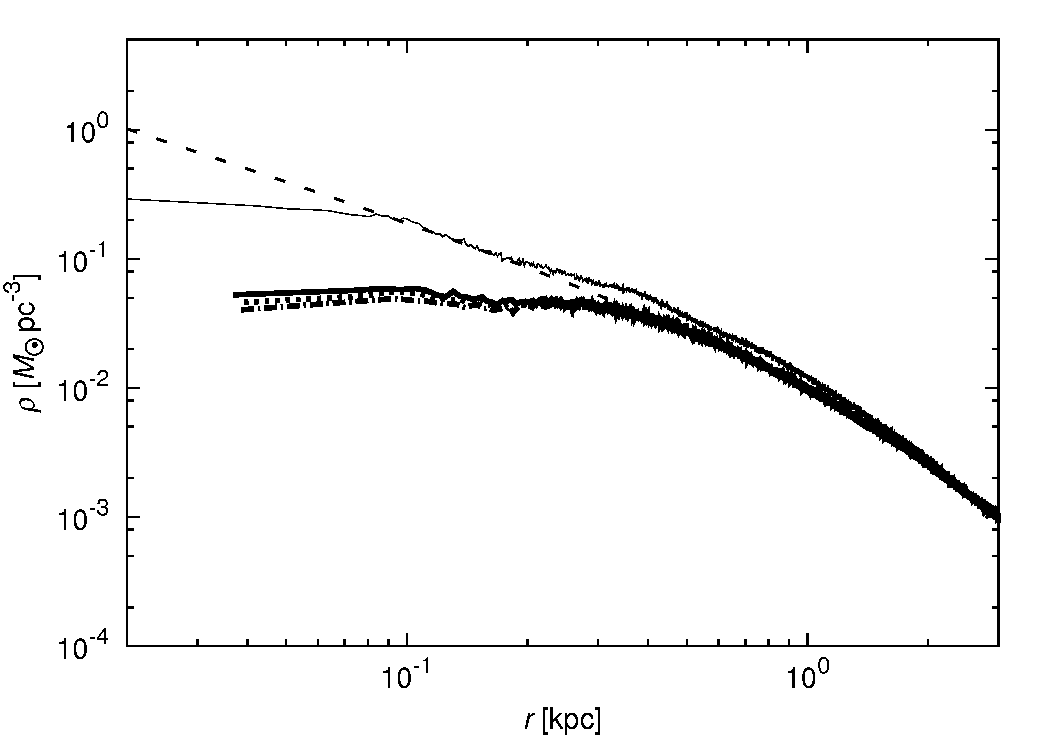}
\caption{
Evolutional process of the density profile of the DM halo (NFW--1) after 10 (thin solid line), 50 (thick solid line), 55 (dotted line), and 60 (dotted--dashed) cycles. The dashed line represents the NFW profile.
\label{many_cycles}
}
\end{figure}

This argument also implies that even if we increase the number of cycles, the structure will be scarcely affected because 
the number of cycles for the excess amount of injected kinetic energy over binding energy, $N_{\rm osc}$, diverges; that is, 
the energy transfer rate, $dK/dt$, approaches zero. We then examine the factor determining such a divergent point of $N_{\rm osc}$, 
or the final core scale. In the initial equilibrium state, particles have velocities less than those escaping from the halo. 
Particles with high velocities decrease with radius and disappear eventually, which indicates that the term in 
Equation (\ref{transfer_rate}), $\frac{d f(v_{\rm 0})}{d v_{\rm 0}}(r) \rightarrow 0$. The slowest velocity for the resonance 
(i.e., the fundamental tone) is given by $v_{\rm 0}(r) = \Omega / k(r)$. Therefore, the divergent point is determined by 
the period of the fundamental tone and the potential profile of the DM halo. The evolutional processes of NFW--1 are shown 
in Figure \ref{many_cycles}, which shows that the growth of the core scale saturates, as expected. 
The core size has grown to 0.20kpc after 60 oscillation cycles. 
This reasonably matches the prediction by Equation (\ref{transfer_rate}) for 60 oscillation cycles, $r_{\rm up}(60) = 0.59{\rm kpc}$.

%
%
%
%

\begin{figure}
\plotone{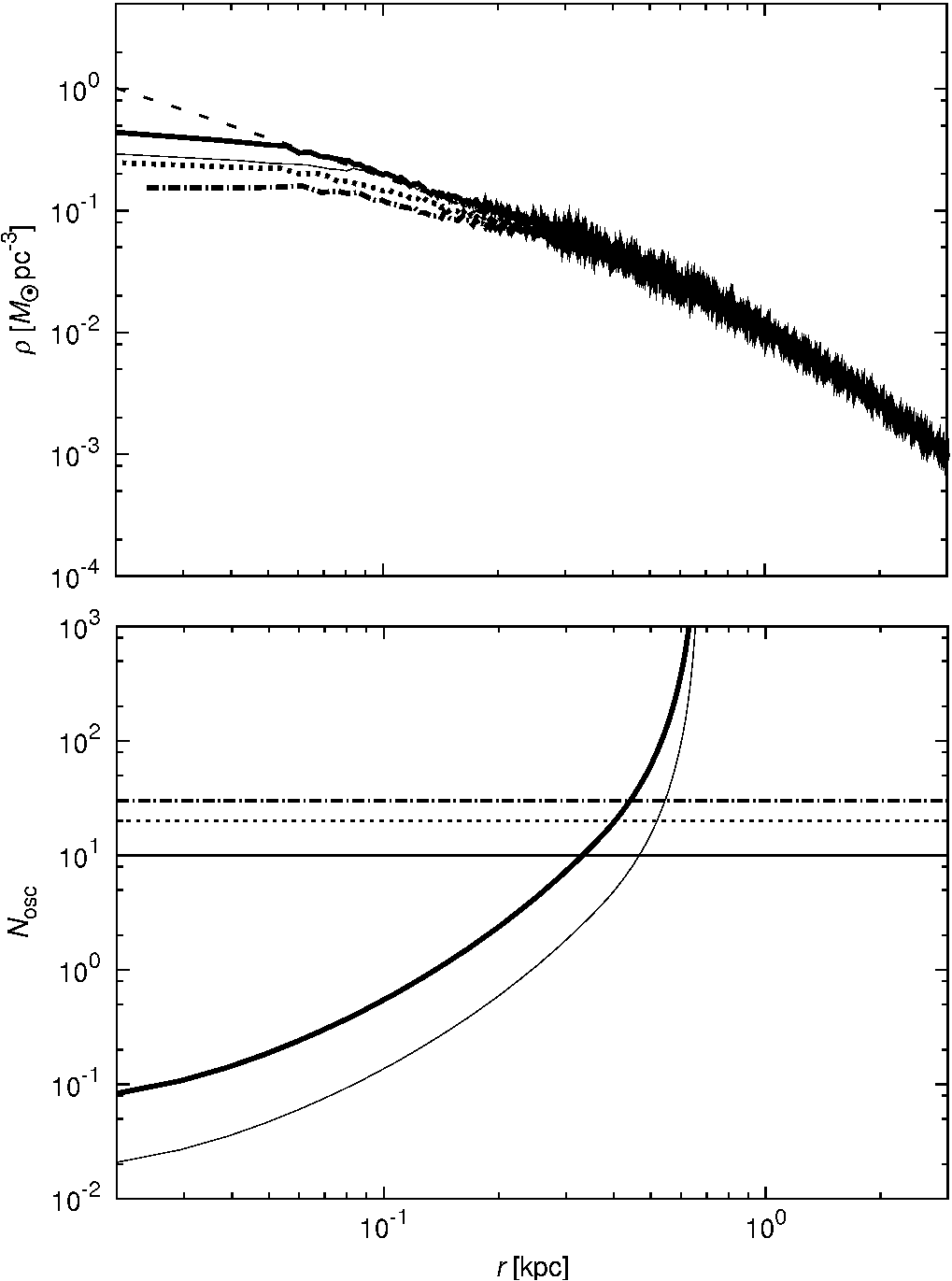}
\caption{
Upper panel shows the evolutional process of the density profile of the DM halo (NFW--8) after 10 (solid line), 20 (dotted line), 
and 30 (dotted--dashed line) cycles. The thin solid and dashed lines express the results of NFW--1 after 10 cycles and the NFW 
profile, respectively. The lower panel shows  $N_{\rm osc}(r)$ as a function of $r$. Thick and thin curves correspond to NFW--8 
and NFW--1, respectively.
\label{half_mass}
}
\end{figure}

By a similar discussion, we argue the dependence on other parameters, $M_{\rm b}, R_{\rm b,max}$, and $R_{\rm b,min}$. In NFW--8, 
the mass of the external potential is set to half, and $T =\tau$. The amplitude of the potential change becomes smaller than 
those in the former runs, and a larger number of oscillation cycles is required to flatten the cusp. We compare $N_{\rm osc}$ 
of NFW--8 with that of NFW--1 in the lower panel of Figure \ref{half_mass}. As expected, $N_{\rm osc}$  becomes larger than NFW--1; 
however, the divergent point is similar to that of NFW--1. The upper panel in the figure shows the evolutional processes of 
the density profile and demonstrates that the number of oscillation cycles required to reach the quasiequilibrium state is 
greater than that in the large amplitude case (NFW--1). However, the final state perfectly matches that of NFW--1.

%
%
%
%

\begin{figure}
\plotone{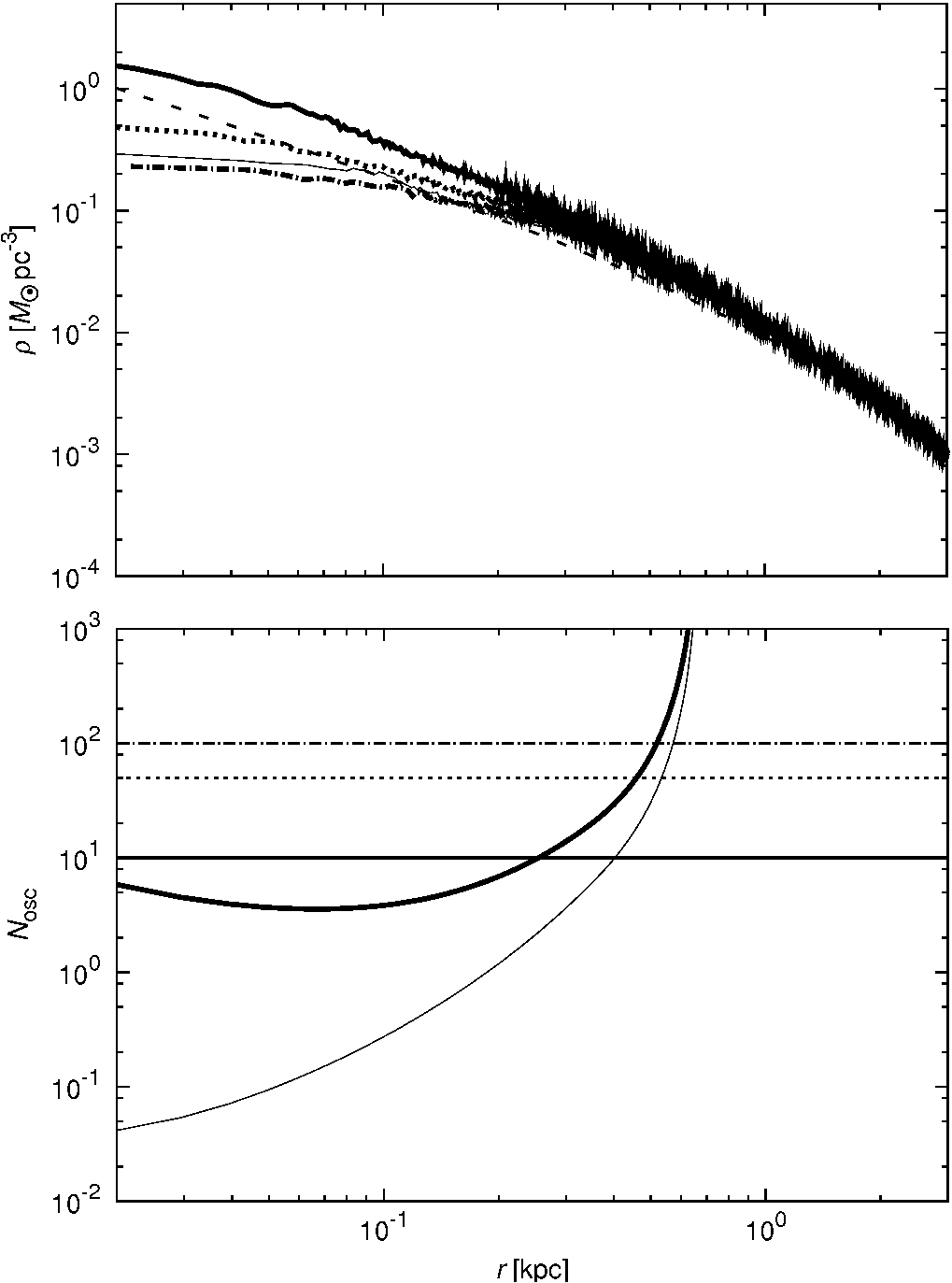}
\caption{
Upper panel shows the evolutional process of the density profile of the DM halo (NFW--9) after 10 (solid line), 50 (dotted line), 
and 100 (dotted--dashed line) cycles. The thin solid and dashed lines express the results of NFW--1 after 10 cycles and the NFW 
profile, respectively. The lower panel shows $N_{\rm osc}(r)$ as a function of $r$. Thick and thin curves correspond to NFW--9 and 
NFW--1, respectively.
\label{large_rbmin}
}
\end{figure}

We set the minimum scale length of the external potential at $R_{\rm b,min}=0.2$ kpc in NFW--9, which is comparable to 
the expected core scale obtained from the former results. The upper and lower panels in Figure \ref{large_rbmin} show 
the evolutional processes of the density profile and $N_{\rm osc}$ of the DM halo, respectively. Similar to the observation 
in NFW--8, $N_{\rm osc}$ becomes larger than NFW--1 because the amplitude of the potential change is smaller, and the divergent 
point of $N_{\rm osc}$ is similar to those observed in NFW--1 and NFW--8. As predicted by the argument of energy transfer rate, 
approximately 100 oscillation cycles are required to flatten the central cusp (upper panel). The DM halo then reaches 
the states of NFW--1 and NFW--8.

%
%
%
%

\begin{figure}
\plotone{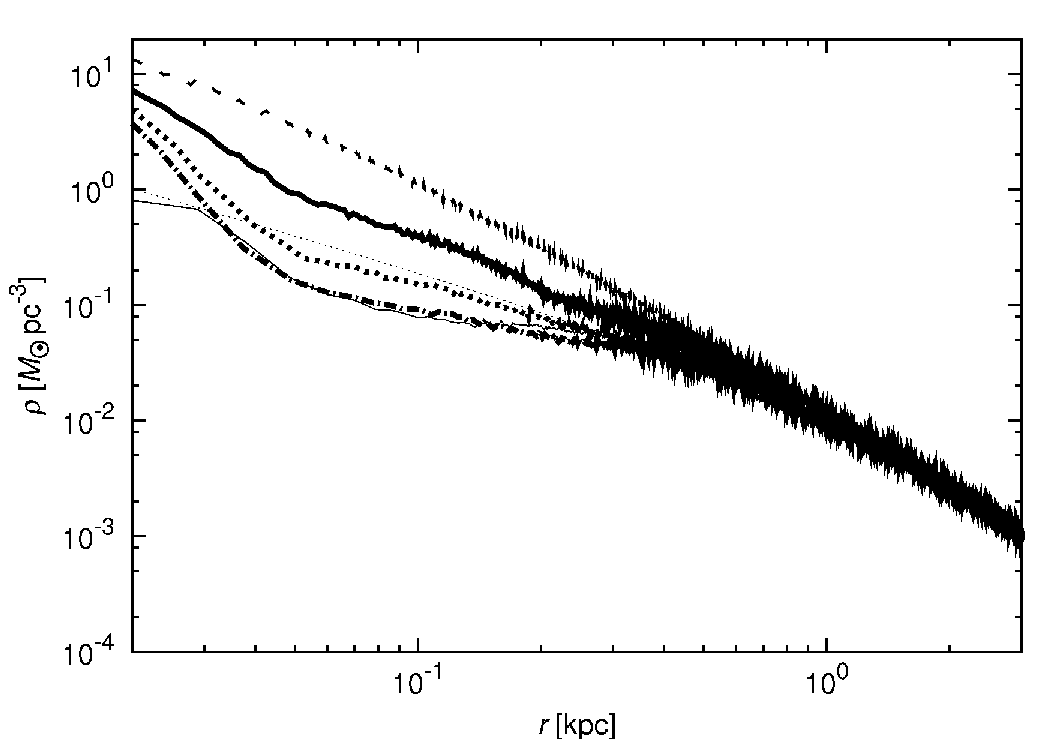}
\caption{
Evolutional process of the density profile of the DM halo (NFW--10) after 10 (solid line), 30 (dotted line), and 50 (dotted-dashed 
line) cycles. The thin solid and thin dashed lines indicate the conditions after 10 cycles of the NFW--11 and NFW profiles, 
respectively. The thick dashed line represents the initial condition, which significantly deviates from the NFW profile because 
the DM halo has relaxed with the static external potential with $R_{\rm b, 0}=0.1$.
\label{small_rbmax}
}
\end{figure}

In the NFW--10 and NFW--11 runs, because we set $R_{\rm b, 0}=0.1$ kpc, the DM halos contract significantly because of the deep 
potential depth of the static external potential and deviate from the NFW profile near the innermost region. $R_{\rm b, max}=0.2$ 
kpc set in NFW--10 is comparable to the core scale predicted by our analytical model and the resultant core scale created 
in the runs of $T =\tau$ (NFW--1, NFW--8, and NFW--9). Figure  \ref{small_rbmax} demonstrates the bump (NFW--10) and 
core (NFW--11) structures appearing near  $r = 0.2 \sim 0.3$ kpc, around which 
core structures have been created in NFW--1, NFW--8 and NFW--9. 
However, even if the number of cycles exceeds 50, the central cusp remains in NFW--10. 
Because the amount of change between $R_{\rm b, min}$ and $R_{\rm b, max}$ is quite small, the external potential acts as an almost static potential. 
In Figure \ref{small_rbmax}, the mass density of NFW--10 decreases gradually in this region, which implies that the cusp-to-core
transition may occur if the number of oscillation cycles is further increased. Because the system deviates from the equilibrium 
NFW model during the relaxation process with the static potential, Widrow's DF is not appropriate for these runs, and thus we 
cannot evaluate $dK/dt$.

Meanwhile, we demonstrate that even if the amplitude of the potential change is small, the resultant density profile is quite 
similar when the oscillation period, $T$, is the same. From these results, we conclude that the period of the oscillation process 
is an important factor in determining the resultant dynamical state governed by the resonance mechanism.

\subsection{Implications}

%
%
%
%

\begin{table*}
\begin{center}
\caption{Comparison with Observational Results}
\begin{tabular}{cccccc}
Galaxy & $r_{\rm core}$ [kpc] & $\rho(r_{\rm core})$ [$M_{\odot} {\rm pc^{-3}}$] & $t_{\rm d}(r_{\rm core})$ [Myr] & $\tau_{\rm SF}$ [Myr] \\
\tableline\tableline
NGC 2366                  & 0.6 & 0.05 & 40 & $\sim {\rm 100}$ \\ 
NGC 6822                  & 1   & 0.03 & 50 & $\sim {\rm 100}$ \\ 
Holmberg I\hspace{-.1em}I & 0.5 & 0.03 & 50 & $\sim {\rm 100}$ \\ 
\tableline
\end{tabular}
\tablecomments{
Approximated values of NGC 2366, NGC 6822, and Holmberg II. $r_{\rm core}$ is the observed core radius, and $\rho(r_{\rm core})$
is the mass density at $r_{\rm core}$. We obtain these values from Weldrake et al. (2003; NGC 6822) and Oh et al. (2008, 2011; 
NGC 2366, Holmberg II). We estimate the dynamical time at $r_{\rm core}$, $t_{\rm d}(r_{\rm core})$ using Equation (\ref{dynamical_time}).
$\tau_{\rm SF}$ is the interval of star formation in these galaxies, which was obtained from McQuinn et al. (2010a, 2010b).
}
\label{obs}
\end{center}
\end{table*}

Gas oscillation is directly correlated with the stellar activity of galaxies. In recent years, highly accurate measurements 
of the SFH and the density profile of nearby dwarf galaxies have been reported. For example, research includes NGC 2366, 
NGC 6822, and Holmberg II (Table \ref{obs}), which have core structures at their centers (Weldrake et al. 2003; Oh et al. 2008, 2011) 
and episodic SFHs (McQuinn et al. 2010a, 2010b). Their dynamical time at the core scale corresponds approximately to 
the intervals of their episodic star formation activities. These galaxies conceivably justify the resonance condition 
predicted by Equation (\ref{core_scale}) at the core scale. 
Yoshii \& Arimoto (1991) studied the color changes in galaxies by using the oscillating star formation history. 
They revealed that oscillatory star formation activity changes the color of galaxies with time and imprints its record 
in the two-color diagram. The oscillation period will be measured more precisely in future observations.

Throughout this paper, we have assumed that the baryon potential oscillates with the constant period and that the baryon 
potential is always active. However, in the real less-massive galaxies, the period changes with time, and the potential 
is shallower because their gas components are gradually ejected by stellar feedback. At the same time, the degree of potential 
change weakens. After ejection of gas components from the galaxies, the central cusp does not recover because this process 
decreases the densities of DM halos (Navarro et al. 1996a; Gnedin \& Zhao 2002; Read \& Gilmore 2005; Ogiya \& Mori 2011; 
Ragone-Figueroa et al. 2012). To explain the physical mechanism of the cusp--core transition of DM halos, we employ an ideal 
model in this study. In subsequent studies, we will examine a more realistic model that includes nonspherical analysis and 
gas dynamics that will more closely examine the correlation between the density structures of DM halos and the SFH of dwarf 
galaxies.

%
%
%
%
\section{Conclusion}
We have studied the dynamical response of DM halos to recurring changes in the gravitational potential of the interstellar 
gas near the centers of the DM halos. A resonance model between the DM particles and density waves excited by the oscillating external potential is 
proposed to understand the physical mechanism of the cusp-core transition of DM halos. We determine that the collisionless 
system effectively gains kinetic energy from the energy transfer driven by the resonance between the DM particles and 
the density waves induced by the oscillation of the gravitational potential of the interstellar gas. The condition for 
the cusp-core transition is such that the oscillation period of the baryon potential is the same as the local dynamical time. 
That is, the resonance condition at a given radius, $r$, is represented by $t_{\rm d}(r_{\rm core}) = T$, and the overtone modes 
are in the resonant states. In addition, the core radius of the DM halo after the cusp--core transition driven by the resonance 
is shown using the conventional mass density profile of DM halos, which is predicted by the cosmological structure formation 
models. Moreover, we verify our analytical model by using $N$-body simulations, whose results validate our resonance model. 
Therefore, we conclude that the energy interchange between the DM particles and the oscillation of the baryon potential driven by the resonance mechanism plays a key 
role in solving the core--cusp problem of DM halos.

\acknowledgments

We thank T. Okamoto, A. Wagner, B. Semelin, and A. Burkert for useful discussions and K. Hasegawa for giving G.O. a fruitful chance. 
We acknowledge the anonymous referee for providing many helpful comments and suggestions. Numerical simulations were performed 
with HA-PACS at the Center for Computational Sciences at the University of Tsukuba. This study was supported in part by JSPS 
Grants-in-Aid for Scientific Research: (A) (21244013), (C) (18540242), and (S) (20224002), Grants-in-Aid for Specially 
Promoted Research by MEXT (16002003), and a Grant-in-Aid for JSPS Fellows (25-1455 G.O.).

%
%
%
%


\begin{thebibliography}{}
\bibitem[Barnes 
\& Hut(1986)]{1986Natur.324..446B} Barnes, J., \& Hut, P.\ 1986, \nat, 324, 446 
\bibitem[Bland-Hawthorn et al.(2011)]{2011EAS....48..397B} Bland-Hawthorn, 
J., Sutherland, R., \& Karlsson, T.\ 2011, EAS Publications Series, 48, 397 
\bibitem[Burkert(1995)]{1995ApJ...447L..25B} Burkert, A.\ 1995, \apjl, 447, 
L25 
\bibitem[de Blok et al.(2001)]{2001AJ....122.2396D} de Blok, W.~J.~G., 
McGaugh, S.~S., \& Rubin, V.~C.\ 2001, \aj, 122, 2396
\bibitem[Del Popolo(2009)]{2009ApJ...698.2093D} Del Popolo, A.\ 2009, \apj, 698, 2093
\bibitem[Diemand et al.(2004)]{2004MNRAS.353..624D} Diemand, J., Moore, B., 
\& Stadel, J.\ 2004, \mnras, 353, 624
\bibitem[Fukushige et al.(2004)]{2004ApJ...606..625F} Fukushige, T., Kawai, A., \& Makino, J.\ 2004, \apj, 606, 625
\bibitem[Fukushige \& Makino(1997)]{1997ApJ...477L...9F} Fukushige, T., \& Makino, J.\ 1997, \apjl, 477, L9
\bibitem[Gentile et al.(2004)]{2004MNRAS.351..903G} Gentile, G., Salucci, 
P., Klein, U., Vergani, D., \& Kalberla, P.\ 2004, \mnras, 351, 903
\bibitem[Gnedin 
\& Zhao(2002)]{2002MNRAS.333..299G} Gnedin, O.~Y., \& Zhao, H.\ 2002, \mnras, 333, 299 
\bibitem[Governato et al.(2010)]{2010Natur.463..203G} Governato, F., Brook, 
C., Mayer, L., et al.\ 2010, \nat, 463, 203
\bibitem[Hernquist(1990)]{1990ApJ...356..359H} Hernquist, L.\ 1990, \apj, 
356, 359
\bibitem[Ikuta 
\& Arimoto(2002)]{2002A&A...391...55I} Ikuta, C., \& Arimoto, N.\ 2002, \aap, 391, 55
\bibitem[Inoue 
\& Saitoh(2011)]{2011MNRAS.418.2527I} Inoue, S., \& Saitoh, T.~R.\ 2011, \mnras, 418, 2527
\bibitem[Ishiyama et al.(2013)]{2013ApJ...767..146I} Ishiyama, T., Rieder, 
S., Makino, J., et al.\ 2013, \apj, 767, 146 
\bibitem[Jing 
\& Suto(2000)]{2000ApJ...529L..69J} Jing, Y.~P., \& Suto, Y.\ 2000, \apjl, 529, L69
\bibitem[Klypin et al.(2001)]{2001ApJ...554..903K} Klypin, A., Kravtsov, 
A.~V., Bullock, J.~S., \& Primack, J.~R.\ 2001, \apj, 554, 903
\bibitem[Komatsu et al.(2009)]{2009ApJS..180..330K} Komatsu, E., Dunkley, 
J., Nolta, M.~R., et al.\ 2009, \apjs, 180, 330 
\bibitem[Komatsu et al.(2011)]{2011ApJS..192...18K} Komatsu, E., Smith, 
K.~M., Dunkley, J., et al.\ 2011, \apjs, 192, 18
\bibitem[Kuzio de Naray et al.(2008)]{2008ApJ...676..920K} Kuzio de Naray, 
R., McGaugh, S.~S., \& de Blok, W.~J.~G.\ 2008, \apj, 676, 920 
\bibitem[Kuzio de Naray et al.(2006)]{2006ApJS..165..461K} Kuzio de Naray, 
R., McGaugh, S.~S., de Blok, W.~J.~G., \& Bosma, A.\ 2006, \apjs, 165, 461
\bibitem[Macci{\`o} et al.(2008)]{2008MNRAS.391.1940M} Macci{\`o}, A.~V., 
Dutton, A.~A., \& van den Bosch, F.~C.\ 2008, \mnras, 391, 1940
\bibitem[Macci{\`o} et al.(2012)]{2012ApJ...744L...9M} Macci{\`o}, A.~V., 
Stinson, G., Brook, C.~B., et al.\ 2012, \apjl, 744, L9
\bibitem[Mac Low 
\& Ferrara(1999)]{1999ApJ...513..142M} Mac Low, M.-M., \& Ferrara, A.\ 1999, \apj, 513, 142
\bibitem[Marcolini et al.(2008)]{2008MNRAS.386.2173M} Marcolini, A., 
D'Ercole, A., Battaglia, G., \& Gibson, B.~K.\ 2008, \mnras, 386, 2173
\bibitem[Marcolini et al.(2006)]{2006MNRAS.371..643M} Marcolini, A., 
D'Ercole, A., Brighenti, F., \& Recchi, S.\ 2006, \mnras, 371, 643
\bibitem[Mashchenko et al.(2006)]{2006Natur.442..539M} Mashchenko, S., 
Couchman, H.~M.~P., \& Wadsley, J.\ 2006, \nat, 442, 539
\bibitem[Mashchenko et al.(2008)]{2008Sci...319..174M} Mashchenko, S., 
Wadsley, J., \& Couchman, H.~M.~P.\ 2008, Science, 319, 174
\bibitem[Mateo(1998)]{1998ARA&A..36..435M} Mateo, M.~L.\ 1998, \araa, 36, 435
\bibitem[McQuinn et al.(2010)]{2010ApJ...721..297M} McQuinn, K.~B.~W., 
Skillman, E.~D., Cannon, J.~M., et al.\ 2010a, \apj, 721, 297
\bibitem[McQuinn et al.(2010)]{2010ApJ...724...49M} McQuinn, K.~B.~W., 
Skillman, E.~D., Cannon, J.~M., et al.\ 2010b, \apj, 724, 49 
\bibitem[Mori et al.(2002)]{2002ApJ...571...40M} Mori, M., Ferrara, A., \& Madau, P.\ 2002, \apj, 571, 40
\bibitem[Mori et al.(1999)]{1999ApJ...511..585M} Mori, M., Yoshii, Y., \& Nomoto, K.\ 1999, \apj, 511, 585
\bibitem[Mori et al.(1997)]{1997ApJ...478L..21M} Mori, M., Yoshii, Y., Tsujimoto, T., \& Nomoto, K.\ 1997, \apjl, 478, L21
\bibitem[Moore(1994)]{1994Natur.370..629M} Moore, B.\ 1994, \nat, 370, 629
\bibitem[Moore et al.(1998)]{1998ApJ...499L...5M} Moore, B., Governato, F., 
Quinn, T., Stadel, J., \& Lake, G.\ 1998, \apjl, 499, L5
\bibitem[Moore et al.(1999)]{1999MNRAS.310.1147M} Moore, B., Quinn, T., 
Governato, F., Stadel, J., \& Lake, G.\ 1999, \mnras, 310, 1147
\bibitem[Nakasato (2012)]{nakasato} Nakasato, N. \ 2012, J. Comp. Sci., 3, 132 
\bibitem[Navarro et al.(1996)]{1996MNRAS.283L..72N} Navarro, J.~F., Eke, 
V.~R., \& Frenk, C.~S.\ 1996a, \mnras, 283, L72
\bibitem[Navarro et al.(1996)]{1996ApJ...462..563N} Navarro, J.~F., Frenk, 
C.~S., \& White, S.~D.~M.\ 1996b, \apj, 462, 563 
\bibitem[Navarro et al.(1997)]{1997ApJ...490..493N} Navarro, J.~F., Frenk, 
C.~S., \& White, S.~D.~M.\ 1997, \apj, 490, 493
\bibitem[Navarro et al.(2004)]{2004MNRAS.349.1039N} Navarro, J.~F., 
Hayashi, E., Power, C., et al.\ 2004, \mnras, 349, 1039
\bibitem[Navarro et al.(2010)]{2010MNRAS.402...21N} Navarro, J.~F., Ludlow, 
A., Springel, V., et al.\ 2010, \mnras, 402, 21
\bibitem[Ogiya \& Mori(2011)]{2011ApJ...736L...2O} Ogiya, G., \& Mori, M.\ 2011, \apjl, 736, L2
\bibitem[Ogiya et al.(2013)]{2013JPhCS.454a2014O} Ogiya, G., Mori, M., 
Miki, Y., Boku, T., 
\& Nakasato, N.\ 2013, JPhCS, 454, 012014 
\bibitem[Oh et al.(2011)]{2011AJ....141..193O} Oh, S.-H., de Blok, 
W.~J.~G., Brinks, E., Walter, F., 
\& Kennicutt, R.~C., Jr.\ 2011, \aj, 141, 193
\bibitem[Oh et al.(2008)]{2008AJ....136.2761O} Oh, S.-H., de Blok, 
W.~J.~G., Walter, F., Brinks, E., 
\& Kennicutt, R.~C., Jr.\ 2008, \aj, 136, 2761
\bibitem[Pontzen 
\& Governato(2012)]{2012MNRAS.421.3464P} Pontzen, A., \& Governato, F.\ 2012, \mnras, 421, 3464
\bibitem[Ragone-Figueroa et al.(2012)]{2012MNRAS.423.3243R} Ragone-Figueroa, C., Granato, G.~L., \& Abadi, M.~G.\ 2012, \mnras, 423, 3243
\bibitem[Read 
\& Gilmore(2005)]{2005MNRAS.356..107R} Read, J.~I., \& Gilmore, G.\ 2005, \mnras, 356, 107
\bibitem[Reed et al.(2005)]{2005MNRAS.357...82R} Reed, D., Governato, F., 
Verde, L., et al.\ 2005, \mnras, 357, 82
\bibitem[Silich 
\& Tenorio-Tagle(2001)]{2001ApJ...552...91S} Silich, S., \& Tenorio-Tagle, G.\ 2001, \apj, 552, 91 
\bibitem[Spekkens et al.(2005)]{2005AJ....129.2119S} Spekkens, K., 
Giovanelli, R., \& Haynes, M.~P.\ 2005, \aj, 129, 2119
\bibitem[Spergel et al.(2007)]{2007ApJS..170..377S} Spergel, D.~N., Bean, 
R., Dor{\'e}, O., et al.\ 2007, \apjs, 170, 377 
\bibitem[Stadel et al.(2009)]{2009MNRAS.398L..21S} Stadel, J., Potter, D., 
Moore, B., et al.\ 2009, \mnras, 398, L21
\bibitem[Stinson et al.(2007)]{2007ApJ...667..170S} Stinson, G.~S., 
Dalcanton, J.~J., Quinn, T., Kaufmann, T., 
\& Wadsley, J.\ 2007, \apj, 667, 170 
\bibitem[Swaters et al.(2003)]{2003ApJ...583..732S} Swaters, R.~A., Madore, 
B.~F., van den Bosch, F.~C., \& Balcells, M.\ 2003, \apj, 583, 732
\bibitem[Teyssier et al.(2013)]{2013MNRAS.429.3068T} Teyssier, R., Pontzen, A., Dubois, Y., \& Read, J.~I.\ 2013, \mnras, 429, 3068 
\bibitem[Tolstoy et 
al.(2009)]{2009ARA&A..47..371T} Tolstoy, E., Hill, V., \& Tosi, M.\ 2009, \araa, 47, 371
\bibitem[Weinberg 
\& Katz(2002)]{2002ApJ...580..627W} Weinberg, M.~D., \& Katz, N.\ 2002, \apj, 580, 627 
\bibitem[Weisz et al.(2011)]{2011ApJ...739....5W} Weisz, D.~R., Dalcanton, 
J.~J., Williams, B.~F., et al.\ 2011, \apj, 739, 5 
\bibitem[Weldrake et al.(2003)]{2003MNRAS.340...12W} Weldrake, D.~T.~F., de 
Blok, W.~J.~G., \& Walter, F.\ 2003, \mnras, 340, 12
\bibitem[Widrow(2000)]{2000ApJS..131...39W} Widrow, L.~M.\ 2000, \apjs, 
131, 39
\bibitem[Yoshii 
\& Arimoto(1987)]{1987A&A...188...13Y} Yoshii, Y., \& Arimoto, N.\ 1987, \aap, 188, 13
\bibitem[Yoshii 
\& Arimoto(1991)]{1991A&A...248...30Y} Yoshii, Y., \& Arimoto, N.\ 1991, \aap, 248, 30
\end{thebibliography}
\end{document}